\documentclass[12pt]{article}

\usepackage{textcomp}
\usepackage{chetnew}
\usepackage{tikz}
\usepackage{amssymb}
\usepackage{comment}

\newcommand{\CH}{\mathcal{H}}
\newcommand{\CQ}{\mathcal{Q}}

\newcommand{\CB}{\mathcal{B}}
\newcommand{\CR}{\mathcal{R}}
\newcommand{\CC}{\mathcal{C}}
\newcommand{\CO}{\mathcal{O}}
\newcommand{\CT}{\mathcal{T}}
\newcommand{\CI}{\mathcal{I}}
\newcommand{\CN}{\mathcal{N}}
\newcommand{\CS}{\mathcal{S}}

\preprint{RU-NHETC-2015-01}

\title{On the Superconformal Index of\\[2mm] Argyres-Douglas~Theories}

\author{Matthew~Buican$^{\diamondsuit}$ and Takahiro~Nishinaka$^{\clubsuit}$
}

\affiliation{NHETC and Department of Physics and Astronomy \\ Rutgers University, Piscataway, NJ 08854, USA
\emails{$^{\diamondsuit}$buican@physics.rutgers.edu,$^{\clubsuit}$nishinaka@physics.rutgers.edu}}

\abstract{We conjecture a closed-form expression for the Schur limit of the superconformal index of two infinite series of Argyres-Douglas (AD) superconformal field theories (SCFTs): the $(A_1,A_{2n-3})$ and the $(A_1,D_{2n})$ theories. While these SCFTs can be realized at special points on the Coulomb branch of certain $\CN=2$ gauge theories, their superconformal $R$ symmetries are emergent, and hence their indices cannot be evaluated by localization. Instead, we construct the $(A_1, A_{2n-3})$ and $(A_1, D_{2n})$ indices by using a relation to two-dimensional $q$-deformed Yang-Mills theory and data from the class $\CS$ construction. Our results generalize the indices derived from the torus partition functions of the two-dimensional chiral algebras associated with the $(A_1, A_3)$ and $(A_1, D_4)$ SCFTs. As checks of our conjectures, we study the consistency of our results with an $S$-duality recently discussed by us in collaboration with Giacomelli and Papageorgakis, we reproduce known Higgs branch relations, we check consistency with a series of renormalization group flows, and we verify that the small $S^1$ limits of our indices reproduce expected Cardy-like behavior. We will discuss the $S^1$ reduction of our indices in a separate paper.
}

\date{May 2015}

\begin{document}
\setcounter{tocdepth}{2}

\maketitle

\toc

\section{Introduction}

Generalized Argyres-Douglas (AD) theories are 4d $\CN=2$ superconformal field theories (SCFTs) with at least one scalar $\CN=2$ chiral generator of non-integer conformal dimension.\foot{By $\CN=2$ chiral generator, we mean a generator of a chiral ring that is annihilated by all the anti-chiral Poincar\'e supercharges of $\CN=2$ SUSY.} In all known examples, these operators can acquire vacuum expectation values (vevs), which then parameterize the Coulomb branch of the theory (indeed, these operators are often referred to as \lq\lq Coulomb branch" operators). AD theories were first discovered at special points on the Coulomb branch of $\CN=2$ supersymmetric gauge theories where BPS states with mutually non-local electro-magnetic charges become simultaneously massless \rcite{Argyres:1995jj, Argyres:1995xn, Eguchi:1996vu}. Since there is no duality frame in which the massless states are only electrically charged, the IR theory is believed to be a strongly coupled SCFT without a Lagrangian description.\foot{This statement is also clear from the fact that at least some of the $\CN=2$ chiral generators have the wrong scaling dimensions and $R$ symmetry quantum numbers to be Casimirs of a weakly coupled gauge group.} These theories were also embedded in string and M-theory in \rcite{Shapere:1999xr, Cecotti:2010fi, Bonelli:2011aa, Xie:2012hs}, which lead to a systematic construction of an infinite number of such SCFTs.

While it has been almost exactly twenty years since AD theories were first discovered, most physical quantities in AD theories (even highly protected observables) still remain to be calculated. The principle reasons that there are so many aspects of AD theories yet to be explored is the lack of a Lagrangian description and the fact that these SCFTs are specified by various subtle scaling limits.\foot{Still, there has been a great deal of recent progress in understanding the $S^1$ reductions of large subclasses of AD theories (see, e.g., \rcite{Xie:2012hs, Boalch:XXXXxx, Boalch:YYYYyy} and references therein), in understanding their BPS spectra (see, e.g., \rcite{Gaiotto:2009hg, Cecotti:2011rv, Alim:2011kw, Alim:2011ae} and references therein), and in understanding properties of their Higgs branches (see, e.g., \rcite{Argyres:2012fu, DelZotto:2014kka} and references therein).}

On the other hand, there is concrete evidence that the simplest-known interacting $\CN=2$ SCFTs are AD theories!  Indeed, we can find some support for this statement by studying the leading behavior of the $a$ conformal anomalies of the AD theories discussed in this paper as a function of their ranks (i.e., the dimensions of their Coulomb branches) and comparing this behavior with the corresponding scaling in other types of theories. For example, taking a Lagrangian gauge theory (with $a=a_{G.T.}$), one of Gaiotto's $T_N$ theories for some particular value of $N=N_0$ \rcite{Gaiotto:2009we} (with $a=a_{T_{N_0}}$), and one of the AD theories described below (with $a=a_{AD}$) all of the same rank, $k\sim N_0^2$, we find
\eqn{
a_{G.T.}\sim~k^2,\ \ \ a_{T_{N_0}}\sim~k^{3\over2}~, \ \ \ a_{AD}\sim k~.
}[aVsRank]
In particular, we see that the total number of degrees of freedom in the AD theory (as measured by $a$) is of the same order as the number of degrees of freedom describing the Coulomb branch! In fact, we will find hints of another much more subtle way in which certain AD theories are simple.

In spite of this simplicity, one important but rather basic observable that has not been computed for AD theories (but has been computed for Lagrangian theories and $T_N$ theories) is the superconformal index.\foot{See, however, \rcite{Buican:2014qla} for a discussion of a particularly simple limit of the index in some AD theories.} This quantity captures the spectrum of certain short multiplets.\foot{We do not distinguish between short and semi-short multiplets.} More precisely, the superconformal index of a four-dimensional $\CN=2$ SCFT is defined as the following refined Witten index, taken with respect to a particular supercharge, $\CQ$, (satisfying $\{\CQ, \CQ^{\dagger}\}=\Delta$) and a (mutually) commuting set of conserved charges \rcite{Romelsberger:2005eg, Kinney:2005ej}
\begin{align}
\text{Tr}_{\mathcal{H}}(-1)^F e^{-\beta \Delta}\, p^{j_1+j_2-r}q^{j_1-j_2-r}t^{R+r}\prod_{i}(x_i)^{f_i}~.
\label{eq:general-index}
\end{align}
In \eqref{eq:general-index}, the trace is taken over the Hilbert space of local operators, $\CH$, the parameters $j_1,j_2$ are $SO(4)$ spins, $R$ is the $SU(2)_R$ Cartan, $r$ is the $U(1)_R\subset U(1)_R\times SU(2)_R$ charge, and the $f_i$ are flavor charges. The fugacities $p$, $q$, $t$, and $x_i$ are complex variables satisfying $|p|$, $|q|$, $|t|$, $|pq/t|<1$, and $|x_i|=1$. By the usual arguments, only operators annihilated by $\CQ$ and $\CQ^{\dagger}$ contribute to the index.\foot{In the conventions of \rcite{Buican:2014qla}, we will choose $\CQ=\tilde\CQ_{2\dot-}$ in this paper (other choices lead to equivalent constructions of the index).} These operators satisfy $\Delta= E-2j_1-2R+r=0$, where $E$ is the scaling dimension. Since the index only counts such short multiplets (modulo combinations of short multiplets that can pair up to form long multiplets), the index is independent of exactly marginal deformations (as long as the theory has a discrete spectrum).

In this paper, we will be primarily interested in the Schur limit of the superconformal index \rcite{Gadde:2011uv}. This limit is defined by taking $t\to q$ in \eqref{eq:general-index}. Since the $p$-dependence drops out in this limit, $p$ is arbitrary, and we find
\begin{align}
\mathcal{I}(q;\vec{x}) = \text{Tr}_{\mathcal{H}}(-1)^F q^{E-R}\prod_{i}(x_i)^{f_i}.
\label{Schurlim}
\end{align}

Alternatively, the superconformal index can be identified as the partition function on $S^1\times S^3$ with background gauge fields (corresponding to the chemical potentials in \eqref{eq:general-index}) turned on \rcite{Romelsberger:2005eg, Kinney:2005ej}. Therefore, in theories with a Lagrangian description, the index can be evaluated by a path integral on $S^1\times S^3$ (via supersymmetric localization or, sometimes, by going to a weak-coupling point on the conformal manifold\footnote{Since the index is independent of marginal couplings, it can be evaluated by a perturbative calculation in the weak-coupling limit.}). However, the superconformal index of AD theories cannot be evaluated in this way because of the reasons mentioned above: the superconformal $R$-symmetry is emergent, and these theories are specified by taking special scaling limits (moreover, even if an AD theory possesses an exactly marginal deformation, there is no free limit on the resulting conformal manifold).

In this paper, we focus on a different route. Guided by the class $\CS$ construction of a certain large class of AD theories and generalizing a 2d/4d correspondence proposed in \rcite{Gadde:2011ik}, we conjecture a closed-form expression for the Schur index of two infinite sets of AD theories: the $(A_1,A_{2n-3})$ and $(A_1,D_{2n})$ theories.
While these theories can be constructed at maximally singular points of pure $SU(2n-2)$ and $SO(4n)$ SYM theory \rcite{Argyres:2012fu} or by putting type IIB string theory on certain Calabi-Yau singularities \rcite{Cecotti:2010fi}, we will find that the most fruitful construction for our purposes comes from the class $\CS$ perspective, i.e., by compactifying a six-dimensional $(2,0)$ theory on a punctured sphere \rcite{Bonelli:2011aa, Xie:2012hs}.\footnote{To be precise, the $(A_1,A_1)$ and $(A_1,D_2)$ theories are {\it not} AD theories but rather are theories of free hypermultiplets. However, they can be embedded in string theory and M-theory exactly in the same way as the other $(A_1,A_{2n-3})$ and $(A_1,D_{2n})$ theories.}

The reason that this latter perspective is so useful is that one can naturally hope to associate a topological quantum field theory (TQFT) with the compactification surface, $\CC$ \rcite{Gadde:2009kb}.
Since the index is invariant under marginal deformations (and these marginal deformations have a geometrical interpretation on $\CC$), it is natural to imagine that it can be computed in this TQFT. Indeed, in the special case of the $(2,0)$ theory on a surface, $\CC$, with \lq\lq regular" punctures, the Schur index of the four-dimensional theory can be thought of as a correlator in two-dimensional $q$-deformed YM theory on $\mathcal{C}$ \rcite{Gadde:2011ik}.\foot{For example, the Schur index of the $T_N$ theory was obtained in this manner.}

Therefore, our strategy in this paper is to generalize the above 2d/4d relation so that it can be applied to the $(A_1,A_{2n-3})$ and $(A_1,D_{2n})$ theories. The main difficulty in our cases of interest is that, as reviewed in section \ref{sec:AD_from_6d}, $\mathcal{C}$ has an ``irregular'' puncture. The interpretation of such a singularity in $q$-deformed YM theory has been unclear. To remedy this situation, we propose just such an interpretation. As a result, our construction leads to a closed-form expression for the Schur index of the $(A_1,A_{2n-3})$ and $(A_1,D_{2n})$ theories.

Let us briefly summarize our conjecture here. As reviewed in section \ref{sec:AD_from_6d}, the flavor symmetries of the $(A_1,A_{2n-3})$ and $(A_1,D_{2n})$ theories are generically $U(1)$ and $U(1)\times SU(2)$, respectively. Let $x$ and $y$ be fugacities for the $U(1)$ and $SU(2)$ flavor subgroups. We conjecture that the Schur indices of the $(A_1,A_{2n-3})$ and $(A_1,D_{2n})$ theories are then given, respectively, by 
\begin{align}
\mathcal{I}_{(A_1,A_{2n-3})}(q;x) &= \CN(q)\sum_{R}[\text{dim}\,R]_q\; \tilde{f}^{(n)}_R(q;x)~,
\label{eq:A1An-index}
\end{align}
and
\begin{align}
\mathcal{I}_{(A_1,D_{2n})}(q;x,y) &= \sum_{R}\tilde{f}_R^{(n)}(q;x)\,f_R(q;y)~,
\label{eq:A1Dn-index}
\end{align}
where $\CN(q)\equiv \prod_{n=2}^\infty(1-q^n)^{-1}$, and $R$ runs over the irreducible representations of $su(2)$. The quantity $\text{dim}\,R$ stands for the dimension of $R$, and $[k]_q \equiv (q^{\frac{k}{2}}-q^{-\frac{k}{2}})/(q^{\frac{1}{2}}-q^{-\frac{1}{2}})$ for any integer $k$. We conjecture that the wave-function factor, $\tilde{f}^{(n)}_R(q;x)$, is given by
\begin{align}
\tilde{f}_R^{(n)}(q;x) &\equiv \frac{q^{nC_2(R)}}{(q;q)_\infty}\text{Tr}_R\left[x^{2J_3} q^{-n(J_3)^2}\right]~,
\label{eq:proposal}
\end{align}
where $(q;q)_\infty = \prod_{k=1}^\infty(1-q^k)$, $J_3$ is a Cartan generator of $su(2)$,\footnote{As usual, we normalize $J_3$ so that the fundamental representation has eigenvalues $J_3=\pm \frac{1}{2}$.} and $C_2(R)$ is the quadratic Casimir invariant of $R$.\footnote{To be explicit, $C_2(R) \equiv ((\text{dim}\,R)^2 - 1)/4$.} The other factor, $f_R(q;x)$, is well-known in the literature on regular singularities, and is given by
\begin{align}
 f_R(q;x)  &\equiv P.E.\left[\frac{q}{1-q}\chi^{su(2)}_{ \text{adj}}(x)\right]\chi_R^{su(2)}(x)~,
\label{eq:trace-regular}
\end{align}
where $\chi^{su(2)}_R(x) = \text{Tr}_{R}\left[x^{2J_3}\right]$ is the $su(2)$ character associated with $R$, and the \lq\lq plethystic exponential" $P.E.[F(q;x_1,\cdots,x_\ell)] \equiv \exp\left(\sum_{k=1}^\infty \frac{F(q^k;x_1^k,\cdots, x_\ell^k)}{k}\right) $ for any function $F$ of the fugacities. As we have mentioned above and will describe in more detail below, the conjectures \eqref{eq:A1An-index} and \eqref{eq:A1Dn-index} are motivated by the 2d/4d relation discovered in \rcite{Gadde:2011ik}.

The plan of this paper is as follows. In section 2, we briefly review the class $\mathcal{S}$ construction of the $(A_1,A_{2n-3})$ and $(A_1,D_{2n})$ theories. In section 3, we further motivate our conjectures \eqref{eq:A1An-index} and \eqref{eq:A1Dn-index}. Then, in the following two sections, we give various consistency checks. In particular, in section 4, we first verify that our conjectures correctly reproduce the known Schur indices of the free $(A_1,A_1)$ and $(A_1,D_2)$ theories. We also check that, in the rank one case, \eqref{eq:A1An-index} and \eqref{eq:A1Dn-index} are consistent with recent connections between two-dimensional chiral algebras and the Schur sector of four-dimensional $\mathcal{N}=2$ SCFTs \rcite{Beem:2013sza}. In addition, we verify that \eqref{eq:A1Dn-index} is consistent with the S-duality found in \rcite{Buican:2014hfa}. In section 5, we perform consistency checks available for general $(A_1,A_{2n-3})$ and $(A_1,D_{2n})$ theories: we study the emergence of known Higgs branch relations, consistency with an intricate set of renormalization group (RG) flows, and expected Cardy-like scaling in the small $S^1$ limit of the index. Finally, we conclude with a discussion of open problems.

\section{Argyres-Douglas theories from 6d $(2,0)$ theories}
\label{sec:AD_from_6d}

A class $\mathcal{S}$ theory is a four-dimensional $\mathcal{N}=2$ supersymmetric field theory, $\mathcal{T}_{\mathcal{C}}$, obtained by compactifying the six-dimensional $(2,0)$ theory with Lie algebra $\mathfrak{g}=A_k$, $D_k$, or $E_k$ on a (punctured) Riemann surface, $\mathcal{C}$ \rcite{Witten:1997sc, Gaiotto:2009we, Gaiotto:2009hg}. Four-dimensional $\mathcal{N}=2$ supersymmetry arises after performing a partial topological twist on $\mathcal{C}$, which breaks the $SO(5)_R$ R-symmetry of the six-dimensional theory down to $SO(2)_R\times SO(3)_R$.
At each puncture on $\mathcal{C}$, a BPS boundary condition is imposed so that four-dimensional supersymmetry is preserved. These BPS boundary conditions correspond to half-BPS co-dimension-two defects extending in the four-dimensional spacetime. Crucially, such BPS defects induce flavor symmetries of $\mathcal{T}_{\mathcal{C}}$. Thus, $\mathcal{T}_{\mathcal{C}}$ is specified by $\mathfrak{g}$, a choice of $\mathcal{C}$, and the BPS defects on $\mathcal{C}$.

In this paper, we focus on the case $\mathfrak{g}=A_1$.  The simplest BPS co-dimension-two defect is then a so-called ``regular'' defect \rcite{Gukov:2006jk, Gaiotto:2009we, Gaiotto:2009hg, Chacaltana:2012zy}, which is characterized by the nearby behavior of a local BPS operator, $\CO$, with $SO(2)_R$ charge $2$ and trivial $SO(3)_R$ quantum numbers\rcite{Gaiotto:2009we, Gaiotto:2009hg}.\footnote{This operator descends from a scalar BPS operator in the 6d theory on ${\bf R}^6$. However, after the topological twist, $\CO$ becomes a differential of degree $2$ on $\mathcal{C}$.} The corresponding vev, $\langle\mathcal{O}\rangle$, parameterizes the Coulomb branch of $\mathcal{T}_{\mathcal{C}}$ and is identified with $\text{Tr}(\varphi^2)$, where $\varphi$ is the Higgs field of an $A_1$ Hitchin system on $\mathcal{C}$ \rcite{Gaiotto:2009hg}.\footnote{The $A_1$ Hitchin system involves an $SU(2)$-bundle, $V$. The Higgs field, $\varphi$, is an $(\text{End}\,V)$-valued meromorphic $(1,0)$-form on $\mathcal{C}$ with possible singularities at the point where a defect is inserted. The moduli space of $\varphi$ with fixed singular behavior corresponds to the Coulomb branch moduli space of the four-dimensional theory, $\mathcal{T}_{\mathcal{C}}$. For example, the Seiberg-Witten curve of $\mathcal{T}_{\mathcal{C}}$ is given by $x^2 = \langle \mathcal{O}\rangle  = \text{Tr}(\varphi^2)$. } At a point where a regular defect is inserted, $\varphi$ has a simple pole.\footnote{If the mass deformations are turned off, $\varphi$ is less singular.} Moreover, the flavor symmetry induced by a regular defect is $SU(2)$. Following the standard terminology in the literature, we call a puncture associated with a regular defect a ``regular puncture.'' If all the punctures on $\mathcal{C}$ are regular punctures, then $\mathcal{T}_{\mathcal{C}}$ is known to be superconformal \rcite{Gaiotto:2009we}. One general property of these SCFTs is that their $\CN=2$ chiral operators always have integer scaling dimensions.

Crucially for us in what follows, there is another class of BPS co-dimension-two defects called ``irregular'' defects \rcite{Witten:2007td, Gaiotto:2009hg, Xie:2012hs}. At a point where an irregular defect is inserted, $\varphi$ has a pole of order $n+1$ (where $n>0$). We call $n$ the ``rank'' of the irregular defect, and it can be any positive half integer. However, we will focus on integer ranks in this paper. The flavor symmetry group induced by an irregular defect is generically $U(1)$. Following the standard terminology, we call a puncture associated with an irregular defect an ``irregular puncture.'' If there is an irregular puncture on $\mathcal{C}$, then $\mathcal{T}_{\mathcal{C}}$ is not always conformal. Indeed, it was shown in \rcite{Bonelli:2011aa, Xie:2012hs} that $\mathcal{T}_{\mathcal{C}}$ is an SCFT if (i) $\mathcal{C}$ is ${\bf P}^1$ with an irregular but no regular puncture, or (ii) $\mathcal{C}$ is ${\bf P}^1$ with an irregular and a regular puncture. In either of these two cases, $\mathcal{T}_{\mathcal{C}}$ generically contains $\CN=2$ chiral operators of {\it non-integer} scaling dimensions \rcite{Bonelli:2011aa, Xie:2012hs, Xie:2013jc}, and therefore is an AD type $\mathcal{N}=2$ SCFT \rcite{Argyres:1995jj,Argyres:1995xn, Eguchi:1996vu}. In this paper we study the superconformal index of two infinite sets of these SCFTs that we will describe in the next sub-section.

\subsection{The $(A_1,A_{2n-3})$ and $(A_1,D_{2n})$ theories}

Suppose that $\mathcal{C}$ is ${\bf P}^1$ with an irregular puncture of rank $n\in {\bf Z}^+$ for $n\geq 2$. Then, the four-dimensional theory, $\mathcal{T}_{\mathcal{C}}$, is an $\mathcal{N}=2$ SCFT called the $(A_1,A_{2n-3})$ theory \rcite{Xie:2012hs, Cecotti:2010fi}. This theory has an interesting moduli space of vacua, with the dimensions of its Coulomb and Higgs branches given by
\begin{align}
\text{dim}_{\bf C}\,\mathcal{M}_{\text{Coulomb}} = n-2~,\ \ \  \text{dim}_{\bf H}\,\mathcal{M}_{\text{Higgs}} = 1~.
\end{align}
The Coulomb branch is parameterized by the vevs of $n-2$ $\CN=2$ chiral operators of scaling dimensions $1+k/n$ for $k=1,2,\cdots,n-2$. The non-integer scaling dimensions imply that the theory does not admit a Lagrangian description for $n\geq 3$. Moreover, the absence of a dimension-two $\CN=2$ chiral operator means that there is no $\mathcal{N}=2$-preserving marginal deformation. On the other hand, the Higgs branch of the theory is known to be the orbifold ${\bf C}^2/{\bf Z}_{n-1}$ \rcite{Argyres:2012fu}. For $n>3$, there is a $U(1)$ flavor symmetry induced by the irregular puncture, but this symmetry is enhanced to $SU(2)$ for $n=2,3$. The fugacity, $x$, in our conjecture \eqref{eq:A1An-index} is associated with this flavor symmetry. 
The conformal anomalies of the theory are \rcite{Xie:2012hs}
\begin{align}\label{acA1A2m3}
a = \frac{12n^2-29n+12}{24n}~,\qquad c=\frac{3n^2-7n+3}{6n}~.
\end{align}
In particular, \eqref{acA1A2m3} is consistent with the fact that the $(A_1, A_1)$ SCFT is a theory of a free hypermultiplet, while the $(A_1, A_3)$ SCFT is the rank-one AD theory obtained at the maximally singular point of the $SU(2)$ gauge theory with two flavors \rcite{Argyres:1995xn}.

Next, consider the case in which $\mathcal{C}$ is a ${\bf P}^1$ with an irregular puncture of rank $n\in{\bf Z}^+$ and a regular puncture. $\mathcal{T}_{\mathcal{C}}$ is now an $\mathcal{N}=2$ SCFT called the $(A_1,D_{2n})$ theory \rcite{Xie:2012hs,Bonelli:2011aa,Cecotti:2010fi}. The dimensions of its Coulomb and Higgs branches are given by
\begin{align}
\text{dim}_{\bf C}\, \mathcal{M}_{\text{Coulomb}} = n-1~,\ \ \  \text{dim}_{\bf H}\,\mathcal{M}_{\text{Higgs}} = 2~.
\end{align}
The $\CN=2$ chiral operators have scaling dimensions $1+k/n$ for $k=1,\cdots,n-1$, which implies that the theory admits no Lagrangian description for $n\geq 2$. There is again no $\mathcal{N}=2$-preserving marginal deformation. The irregular puncture induces a $U(1)$ flavor subgroup, while the regular one induces $SU(2)$. For $n\ge3$, the flavor symmetry of the theory is indeed $U(1)\times SU(2)$. However, for $n=2$ the symmetry is enhanced to $SU(3)$, while for $n=1$ it is enhanced to $Sp(2)$. The fugacities, $x$ and $y$, in our conjecture \eqref{eq:A1Dn-index} are associated with the $U(1)$ and $SU(2)$ subgroup of the flavor symmetry respectively. The flavor central charge for the $SU(2)$ subgroup is $k_{SU(2)}=\frac{4n-2}{n}$ \rcite{Aharony:2007dj, Shapere:2008zf}.\footnote{Here we use the normalization in which a fundamental hypermultiplet contributes $k_{SU(2)} = 2$.} The conformal anomalies of the theory are \rcite{Xie:2012hs}
\begin{align}\label{acA1D2n}
a = \frac{6n-5}{12}~,\ \ \ c= \frac{3n-2}{6}~.
\end{align}
In particular, \eqref{acA1D2n} is consistent with the fact that the $(A_1,D_2)$ SCFT is the theory of two free hypermultiplets, and the $(A_1,D_4)$ SCFT is the rank-one AD theory obtained at the maximally singular point of the $SU(2)$ gauge theory with three flavors \rcite{Argyres:1995xn}. 

\section{Motivating our conjectures for the superconformal index}
\label{subsec:motivation}

The main claims of this paper are the conjectures \eqref{eq:A1An-index} and \eqref{eq:A1Dn-index} for the Schur limit of the superconformal index of the $(A_1,A_{2n-3})$ and $(A_1,D_{2n})$ theories described above. For ease of reference, we reproduce these conjectures below. We have
\begin{align}
\mathcal{I}_{(A_1,A_{2n-3})}(q;x) &= \mathcal{N}(q)\sum_{R}[\text{dim}\,R]_q\; \tilde{f}^{(n)}_R(q;x)~, \ \ \ \tilde{f}_R^{(n)}(q;x) &\equiv \frac{q^{nC_2(R)}}{(q;q)_\infty}\text{Tr}_R\left[x^{2J_3} q^{-n(J_3)^2}\right]~,
\label{eq:A1An-indexii}
\end{align}
where $\CN(q)\equiv \prod_{n=2}^\infty(1-q^n)^{-1}$, ${\rm dim}\ R$ is the dimension of the $su(2)$ representation, $R$, and $[k]_q\equiv(q^{k\over2}-q^{-{k\over2}})/(q^{1\over2}-q^{-{1\over2}})$. On the other hand, for the $(A_1, D_{2n})$ theories, we have
\begin{align}
\mathcal{I}_{(A_1,D_{2n})}(q;x,y) &= \sum_{R}\tilde{f}_R^{(n)}(q;x)\,f_R(q;y)~, \ \ \  f_R(q;x)  &\equiv P.E.\left[\frac{q}{1-q}\chi^{su(2)}_{ \text{adj}}(x)\right]\chi_R^{su(2)}(x)~.
\label{eq:A1Dn-indexii}
\end{align}
In this section, we would like to motivate these two conjectures.

Our task is made somewhat more difficult by the fact that the AD theories we study lack a Lagrangian description. Therefore, as mentioned in the introduction, we cannot compute their indices using the machinery of supersymmetric localization. However, as described above, there is a potential way out for theories of class $\CS$. Indeed, since the index is invariant under exactly marginal deformations, and since these exactly marginal deformations have geometrical interpretations on the Riemann surface, $\CC$, used to define the class $\CS$ theory, $\CT_{\CC}$, it is natural to expect that the index can be interpreted as a quantity in a TQFT on $\CC$. Indeed, for a choice of $\mathcal{C}$ {\it without} irregular punctures, the superconformal index of $\mathcal{T}_{\mathcal{C}}$ is known to be equivalent to a correlator of a two-dimensional TQFT on $\mathcal{C}$ \rcite{Gadde:2009kb} (see also \rcite{Gadde:2011ik, Gadde:2011uv, Gaiotto:2012xa, Mekareeya:2012tn, Fukuda:2012jr, Tachikawa:2015iba}). Therefore, we need to generalize this index/TQFT relation to the case in which we have an irregular puncture.

In order to understand how to proceed, let us first review in greater detail the case in which $\mathcal{C}$ is a Riemann surface with genus $g$ and $m$ {\it regular} punctures. In this case, $\mathcal{T}_{\mathcal{C}}$ is an $\mathcal{N}=2$ SCFT with at least $SU(2)^m$ flavor symmetry,\footnote{Recall that we are focusing on $\mathfrak{g}=A_1$.} whose Schur index is known to be \rcite{Gadde:2011ik}
\begin{align}
\mathcal{I}_{\mathcal{T}_{\mathcal{C}}}(q;\vec{x}) = \big[\mathcal{N}(q)\big]^{2-2g-m} \sum_{R}\left([\text{dim}\, R]_q\right)^{2-2g-m}\prod_{k=1}^{m}f_R(q;x_k)~.
\label{eq:regular}
\end{align}
Here $R$ runs over the irreducible representations of $su(2)$, and the factors $\mathcal{N}(q),\,[k]_q$ and $f_R(q;x)$ are given above. The variable, $x_k$, is the fugacity for the $SU(2)$ flavor subgroup induced by the $k$-th regular puncture.
Up to a prefactor, the right-hand side of \eqref{eq:regular} is precisely the $m$-point function of $q$-deformed two-dimensional $SU(2)$ YM theory on $\mathcal{C}$ in the zero-area limit \rcite{Gadde:2011ik}.\footnote{If the area of $\mathcal{C}$ is non-zero, the 4d theory becomes an $\mathcal{N}=2$ sigma model. The relation to two-dimensional YM theory has a natural generalization to this case \rcite{Tachikawa:2012wi}.} Furthermore, as reviewed in appendix \ref{app:qYM}, the sum over $R$ represents the sum over intermediate states, while $f_R(q;x_k)$ is the inner product of the intermediate state $|R\rangle$ and an external state $|x_k\rangle$ (i.e., it is a wave function)
\begin{align}
\langle R| x_k\rangle = f_R(q;x_k)~.
\label{eq:inner-regular}
\end{align}
This discussion implies that each {\it regular} puncture corresponds to an external state, $|x_k\rangle$, of $q$-deformed YM theory. The state, $|x_k\rangle$, is associated with a fixed holonomy, $x_k$, of the $SU(2)$ gauge field around the puncture. 
 Note that \eqref{eq:regular} is independent of the ordering of the $m$ punctures, which reflects the fact that the two-dimensional YM theory in the zero-area limit is a TQFT. 

Since the superconformal index is the $S^1\times S^3$ partition function of the four-dimensional theory with background gauge fields turned on, this connection to two-dimensional YM theory can be regarded as an $S^1\times S^3$ version of the AGT relation \rcite{Alday:2009aq}. Indeed, recall that the AGT relation maps the $S^4$ partition function of $\mathcal{T}_{\mathcal{C}}$ to a correlator of Liouville theory on $\mathcal{C}$. As we alluded to above, the emergence of a TQFT instead of Liouville theory in the case of the $S^3\times S^1$ partition function reflects the fact that, unlike the $S^4$ partition function, the $S^1\times S^3$ partition function is independent of four-dimensional marginal couplings, which are now encoded in the loci of the punctures on $\mathcal{C}$ \rcite{Gadde:2009kb}.

While it has been unclear whether this index/TQFT relation can be generalized to AD theories---and the $(A_1,A_{2n-3})$ and $(A_1,D_{2n})$ theories in particular---there is actually a natural generalization of the AGT relation to these theories \rcite{Gaiotto:2009ma, Bonelli:2011aa, Gaiotto:2012sf, Kanno:2013vi}. A key observation in these works is that, while a regular puncture corresponds to a Virasoro primary state, an irregular puncture corresponds to a coherent state in the Verma module. Such a state is a highly non-trivial linear combination of the primary and an infinite number of descendants.

Given the success of the generalized AGT relation, it is natural to expect that a similar generalization is possible for the superconformal index. In particular, we expect that there exists a state, $|x;n\rangle$, in the $q$-deformed two-dimensional YM theory which corresponds to an irregular puncture of rank $n$ with $U(1)$ flavor fugacity, $x$. Since the Hilbert space of two-dimensional YM theory is spanned by the states $|R\rangle$, we see that $|x;n\rangle$ is uniquely determined by specifying the wave function
\begin{align}
\langle R| x;n\rangle \equiv \tilde{f}^{(n)}_R(q;x)~,
\label{eq:inner-irregular}
\end{align}
for all the irreducible representations, $R$, of $su(2)$. Once this inner product is given, our conjectures \eqref{eq:A1An-indexii} and \eqref{eq:A1Dn-indexii} are natural generalizations of \eqref{eq:regular} to the $(A_1,A_{2n-3})$ and $(A_1,D_{2n})$ theories.

Indeed, for the $(A_1,A_{2n-3})$ theory, the Riemann surface $\mathcal{C}$ is a ${\bf P}^1$ with an irregular puncture of rank $n$. Therefore, we set $g=0$ and $m=1$ in \eqref{eq:regular}, and then replace $f_R(q;x)$ with $\tilde{f}_R^{(n)}(q;x)$. This leads to the general form of our conjecture \eqref{eq:A1An-indexii}. On the other hand, for the $(A_1,D_{2n})$ theory, $\mathcal{C}$ is a ${\bf P}^1$ with an irregular puncture of rank $n$ and a regular puncture. The irregular puncture corresponds to $|x;n\rangle$ while the regular one corresponds to $|y\rangle$. Setting $g=0$ and $m=2$ in \eqref{eq:regular} and replacing one of the two $f_R$ with $\tilde{f}_R^{(n)}$ leads to the general form of our conjecture \eqref{eq:A1Dn-indexii}.

Therefore, the only non-trivial task is to determine the inner product \eqref{eq:inner-irregular}. We conjecture that the inner product, $\tilde{f}^{(n)}_R(q;x)$, is of the form given in \eqref{eq:A1An-indexii}. Let us briefly discuss why this proposal is natural.

To that end, first note that, in contrast to the expression for $f_R$, the expression for $\tilde f_R^{(n)}$ is not written in terms of the characters of $su(2)$. In particular, when $\tilde{f}_R^{(n)}$ is expanded in powers of $q$, the expansion coefficients are not given by $su(2)$ characters. This property is consistent with the fact that, unlike a regular puncture, an irregular puncture generically only gives rise to a $U(1)$ flavor symmetry. Indeed, the $x$ appearing in $\tilde{f}_R^{(n)}(q;x)$ is a $U(1)$ (not $SU(2)$) flavor fugacity. Nevertheless, $(q;q)_\infty \tilde{f}^{(n)}_{R}(q;x)$ reduces to a character of $su(2)$ in the limit $q\to 1$:
\begin{align}
(q;q)_\infty \tilde{f}^{(n)}_{R}(q;x)\rightarrow \chi_{R}^{su(2)}(x)~.
\label{eq:correspondance}
\end{align}
Therefore, $(q;q)_\infty \tilde{f}_R^{(n)}(q;x)$ can be regarded as a natural deformation of $\chi_R^{su(2)}(x)$ under the modification of the flavor symmetry from $SU(2)$ to $U(1)$.
Furthermore, the prefactor $[(q;q)_\infty]^{-1} = P.E.\left[\frac{q}{1-q}\right]$ is a natural generalization of the prefactor, $P.E.\Big[\frac{q}{1-q}\chi_{\text{adj}}^{su(2)}(x)\Big]$, that appears in $f_R(q;x)$, since the character for the adjoint representation of $u(1)$ is $1$.

Another nice property of our expression for $\tilde{f}_R^{(n)}(q;x)$ is that it guarantees that the indices \eqref{eq:A1An-indexii} and \eqref{eq:A1Dn-indexii} are both non-singular at $q=0$. Indeed, this fact follows from
\begin{align}
\tilde{f}_R^{(n)}(q;x) = \mathcal{O}(q^{n(\text{dim} R\,-\,1)/2})~,\ \ \ [\text{dim}\, R]_q = \mathcal{O}(q^{-(\text{dim} R\,-\,1)/2})~.
\end{align} 
This behavior is consistent with the fact that any Schur operator (i.e., an operator contributing to the Schur index) in any unitary four-dimensional $\mathcal{N}=2$ SCFT has $E-R\geq 0$.
In the next two sections, we will give various non-trivial consistency checks of our conjectures.

\section{Consistency with known results for lower-rank cases}
\label{sec:chiral_alg}

In this section, we check that our conjectures are consistent with several indices that are already known (or, in some cases, strongly believed to be known). 

\subsection{The $(A_1,A_1)$ and $(A_1,D_2)$ theories}

The $(A_1,A_1)$ and $(A_1,D_2)$ SCFTs are theories of free hypermultiplets whose indices can be evaluated using their Lagrangian description \rcite{Romelsberger:2005eg, Kinney:2005ej}. It is therefore crucial to check that our formulas reproduce these computations.

The $(A_1,A_1)$ theory is the theory of a single free hypermultiplet, and so its Schur index is given by
\begin{align}
  \prod_{k=0}^\infty \frac{1}{(1-q^{k+\frac{1}{2}}x)(1-q^{k+\frac{1}{2}}x^{-1})}~.
\label{eq:A1A1-known}
\end{align}
Here $x$ is the fugacity for the flavor $Sp(1)\simeq SU(2)$ symmetry such that the character of the fundamental representation of $SU(2)$ is $x + x^{-1}$.
On the other hand, our conjecture \eqref{eq:A1An-indexii} implies
\begin{align}
\mathcal{I}_{(A_1,A_1)}(q;x) = \mathcal{N}(q)\sum_R\, [\text{dim}\,R]_q \,\tilde{f}_R^{(2)}(q;x)~.
\label{eq:A1A1-ours}
\end{align}

We have performed various checks of the equivalence of \eqref{eq:A1A1-known} and \eqref{eq:A1A1-ours}. For example, we have checked this statement to high order perturbatively in $q$. We can also compare the analytic behaviors of \eqref{eq:A1A1-known} and \eqref{eq:A1A1-ours} as functions of $x$. 
Indeed, the expression \eqref{eq:A1A1-ours} has simple poles at $x^\pm = q^{k+\frac{1}{2}}$ for all $k\ge0$ and is regular at all the other points in the complex $x$-plane (see the discussion in sub-section \ref{subsec:Higgsing} for more details). Moreover, the residue at $x^\pm =q^{k+\frac{1}{2}}$ is $\pm (-1)^k q^{\frac{k(k+1)}{2} \pm (k +\frac{1}{2})}\big[(q;q)_\infty\big]^{-2}$. These poles and residues coincide with those of \eqref{eq:A1A1-known}. This evidence strongly suggests that \eqref{eq:A1A1-known} and \eqref{eq:A1A1-ours} are equivalent. Also, note that the manifest $U(1)$ flavor symmetry in \eqref{eq:A1A1-ours} is now enhanced to $Sp(1)\simeq SU(2)$.

The next simplest example is the $(A_1,D_2)$ theory, which is the theory of two free hypermulitplets. Its Schur index is
\begin{align}
 \prod_{k=0}^\infty\prod_{s_1,s_2=\pm 1}\frac{1}{(1-q^{k+\frac{1}{2}} x^{s_1} y^{s_2})}~,
\label{eq:A1D2-index1}
\end{align}
where $x$ and $y$ are fugacities for the flavor symmetry subgroup $SU(2)\times SU(2)\subset Sp(2)$. On the other hand, our conjecture \eqref{eq:A1Dn-indexii} implies
\begin{align}
\mathcal{I}_{(A_1,D_2)}(q;x,y) = \sum_{R}\tilde{f}^{(1)}_{R}(q;x)f_{R}(q;y)~.
\label{eq:A1D2-index2}
\end{align}
Again, we have checked the equivalence of \eqref{eq:A1D2-index1} and \eqref{eq:A1D2-index2} to high order perturbatively in $q$. We have also checked that they share the same analytic behavior as functions of $x$ and $y$. Note that the manifest $SU(2)\times U(1)$ flavor symmetry of \eqref{eq:A1D2-index2} is therefore appropriately enhanced.

While the $(A_1, A_1)$ and $(A_1, D_2)$ theories are particularly simple, the above agreement is highly non-trivial since our conjectures do not rely on Lagrangian descriptions of these theories.

\subsection{The $(A_1,A_3)$ and $(A_1,D_4)$ theories and two-dimensional chiral algebras}
\label{subsec:2d-chiral}

The authors of \rcite{Beem:2013sza} showed that the Schur operators (again, these are the operators contributing to the Schur limit of the index) of any four-dimensional $\mathcal{N}=2$ SCFT, $\mathcal{T}$, with flavor symmetry, $G_F$, map to a two-dimensional chiral algebra, $\chi_{\mathcal{T}}$, containing a Virasoro sub-algebra and an affine Kac-Moody (AKM) sub-algebra associated with $G_F$. In this sub-section, we will check the consistency of our conjectures \eqref{eq:A1An-indexii} and \eqref{eq:A1Dn-indexii} with this observation. We will primarily focus on the cases of the $(A_1,A_3)$ and $(A_1,D_4)$ theories (however, we will present some comments on the more general cases in sub-section \ref{subsubsec:general-AD}).

The essential elements of \rcite{Beem:2013sza} that we will need are the following. First, the basic two-dimensional quantity that we can compare with the Schur index is the torus partition function of the chiral algebra. In particular, a basic entry in the dictionary of \rcite{Beem:2013sza} states that for a given four-dimensional theory, $\CT$, and a corresponding two-dimensional chiral algebra, $\chi_{\CT}$, we have
\eqn{
{\rm Tr}_{\chi_{\CT}}(-1)^Fq^{L_0}\prod_i(x_i)^{f_i}=\CI_{\CT}(q;x_i)~,
}[torusIndex]
where the quantity on the RHS of \eqref{torusIndex} is the Schur index of $\CT$, and the quantity on the LHS is the (graded) trace over the Hilbert space of the corresponding chiral algebra weighted by the same fugacities as in the four-dimensional theory but with the four-dimensional generator, $E-R$, replaced by the Virasoro generator, $L_0$. Therefore, once we identify the two-dimensional chiral algebra, $\chi_{\mathcal{T}}$, we can immediately compute the Schur index of the four-dimensional theory, $\mathcal{T}$, through the relation \eqref{torusIndex}.

To identify $\chi_{\mathcal{T}}$, it is useful to know how the most basic universal observables of the two-dimensional theory---the central charge, $c_{2d}$, and (when there is a flavor symmetry) the AKM level(s), $k_{2d}$---map to four dimensions. According to the discussion in \rcite{Beem:2013sza}, these quantities are determined in terms of the four-dimensional central charge, $c_{4d}$, and four-dimensional flavor anomaly, $k_{4d}$, via
\eqn{
c_{2d}=-12c_{4d}~, \ \ \ k_{2d}=-{1\over2}k_{4d}~.
}[2d4dobs]
In particular, it follows from \eqref{2d4dobs} that the two-dimensional theory is non-unitary if the four-dimensional theory is unitary. Moreover, since flavor anomalies of AD theories are generically non-integer, it follows that the AKM level will also generically be non-integer. 

Let us now focus on the cases of the $(A_1, A_3)$ and $(A_1, D_4)$ SCFTs, where the corresponding chiral algebra construction is most obvious. We will come back to general AD theories later in sub-section \ref{subsubsec:general-AD}. From \eqref{acA1A2m3}, \eqref{acA1D2n} (including the text above this equation), and \eqref{2d4dobs}, we see that, in the case of the $(A_1,A_3)$ theory,
\eqn{
c_{2d}^{(A_1, A_3)}=-6~, \ \ \ k_{2d}^{(A_1, A_3)}=-{4\over3}~,
}[obsA1A3]
while in the case of the $(A_1, D_4)$ theory, we have
\eqn{
c_{2d}^{(A_1, D_4)}=-8~, \ \ \ k_{2d}^{(A_1, D_4)}=-{3\over2}~.
}[obsA1D4]
In particular, we see that the chiral algebras, $\chi_{(A_1 A_3)}$ and $\chi_{(A_1, D_4)}$, contain, respectively, an $su(2)$ AKM sub-algebra at level $-4/3$, $\widehat{su}(2)_{-{4\over3}}$, and an $su(3)$ AKM sub-algebra at level $-3/2$, $\widehat{su}(3)_{-{3\over2}}$. Moreover, the values of $c_{2d}$ and $k_{2d}$ in both cases are precisely the values for which the central charges of the Sugawara stress tensors associated with the AKM sub-algebras coincide with the central charges of the Virasoro algebras. In fact, as the authors of \rcite{Beem:2013sza} showed, this phenomenon follows from the saturation of certain four-dimensional unitarity bounds in all known rank-one SCFTs.\foot{These are the SCFTs realized on a D3-brane probing an F-theory singularity of Kodaira type \rcite{Sen:1996vd, Banks:1996nj, Dasgupta:1996ij, Minahan:1996fg, Minahan:1996cj, Aharony:1998xz}.} Furthermore, these authors showed that for theories saturating these unitarity bounds, the two dimensional stress tensor {\it is} the Sugawara stress tensor.

As a result, a minimal conjecture for the chiral algebras of the $(A_1, A_3)$ and $(A_1, D_4)$ SCFTs is that they are simply $\widehat{su}(2)_{-{4\over3}}$ and $\widehat{su}(3)_{-{3\over2}}$ respectively. If this conjecture is correct, then the only two-dimensional operators we need to include are in the vacuum module of the two AKM algebras. Indeed, the authors of \rcite{Beem:2013sza} used similar reasoning in the case of the $T_3$ and $SU(2)$ with $N_f=4$ rank-one theories, where the analogous statements were more explicitly investigated.

\subsubsection{The $(A_1,A_3)$ theory}
\begin{table}
\begin{center}
\begin{tabular}{cl}
\hline\hline
level & $su(2)$ representations and their multiplicities \\ 
\hline
$0$ & ${\bf 1}$ \\
$1$ & ${\bf 3}$ \\
$2$ & ${\bf 1},\;\,{\bf 3},\;\, {\bf 5}$ \\
$3$ & ${\bf 1},\;\, 2\times {\bf 3},\;\, {\bf 5},\;\, {\bf 7}$\\
$4$ & $2 \times {\bf 1},\;\, 3\times {\bf 3},\;\, 3\times {\bf 5},\;\, {\bf 7},\;\, {\bf 9}$\\
\hline
\end{tabular}
\caption{The multiplicities of $su(2)$ representations at the first five levels in \eqref{eq:A1A3-index}.}
\label{table:A1A3-multiplicities}
\end{center}
\end{table}
Given our discussion above, we should compare the torus partition function (or character) of the vacuum module of $\widehat{su}(2)_{-{4\over3}}$, $\CH_0^{[-{4\over3},0]}$, with our conjecture for the Schur index. Note that the vacuum module has highest weight, $\lambda=[-\frac{4}{3},0]$.\footnote{Here $[\lambda_0,\lambda_1]$ are the Dynkin labels, namely, $\langle \lambda,\alpha_i^\vee\rangle = \lambda_i$. The torus partition function (or character) is normalized so that the highest-weight state contributes $1$, which coincides with the condition that the unit operator of the four-dimensional theory, $\mathcal{T}_{\mathcal{C}}$, contributes $1$ to the superconformal index.}  In particular, we should have
\begin{align}
\mathcal{I}_{(A_1,A_3)}(q;z) = {\rm Tr}_{\CH_0^{[-{4\over3},0]}}(-1)^F q^{L_0}z^{f}\equiv e^{-\lambda}\text{ch}\, L(\lambda)~,
\label{eq:A1A3-character}
\end{align}
where we use \lq\lq$\equiv$" to underline the fact that the torus partition function is equivalent to the character. This latter terminology is used in the mathematics literature that will be relevant to us below.
The fugacities are identified as $q = e^{-\delta}$ and $z = e^{-\frac{1}{2}\alpha_1}$, where the notation is summarized in appendix \ref{app:character}. 

The right-hand side of \eqref{eq:A1A3-character} can be evaluated order-by-order in $q$ via the character formula proven in \rcite{Kac-Wakimoto} (see appendix \ref{app:character} for more details). The result is written as
\begin{align}
e^{-\lambda}\text{ch}\, L(\lambda) =\; & 1 + \chi_{\bf 3}^{su(2)}(z)q + \big[1 + \chi_{\bf 3}^{su(2)}(z) +\chi_{\bf 5}^{su(2)}(z)\big]q^2  + \cdots~,
\label{eq:A1A3-index}
\end{align}
where $\chi_{\bf n}^{su(2)}(z) = (z^{n}-z^{-n})/(z-z^{-1})$ are characters of $su(2)$.
Note that the $SU(2)$ flavor symmetry of the $(A_1,A_3)$ theory is manifest in this expression. The multiplicities of the $su(2)$ representations up to $\mathcal{O}(q^4)$ are summarized in table \ref{table:A1A3-multiplicities}.

On the other hand, our conjecture \eqref{eq:A1An-index} implies
\begin{align}
\mathcal{I}_{(A_1,A_3)}(q;x) = \mathcal{N}(q)\sum_{R}[\text{dim}\,R]_q\tilde{f}^{(3)}_{R}(q;x)~.
\label{eq:conjecture3}
\end{align}
We have checked that \eqref{eq:A1A3-index} and \eqref{eq:conjecture3} agree up to very high order perturbatively in $q$ under the identification $z=\sqrt{x}$.\footnote{Note that $z$ and $x$ are fugacities for $SU(2)$ and $U(1)$, respectively. The relation $z=\sqrt{x}$ simply means that the $SU(2)$ spin associated with $z$ is identified as the $U(1)$ charge associated with $x$.} This result is a highly non-trivial consistency check of our conjecture. In particular, note that this agreement implies that the manifest $U(1)$ flavor symmetry in \eqref{eq:conjecture3} is enhanced to $SU(2)$.

\subsubsection{The $(A_1,D_4)$ theory}

\begin{table}
\begin{center}
\begin{tabular}{cl}
\hline\hline
level & $su(3)$ representations and their multiplicities \\
\hline
$0$ & ${\bf 1}$ \\
$1$ & ${\bf 8}$ \\
$2$ & ${\bf 1},\;\,{\bf 8},\;\, {\bf 27}$ \\
$3$ & ${\bf 1},\;\,2\times {\bf 8},\;\, {\bf 10},\;\, \overline{\bf 10},\;\, {\bf 27},\;\, {\bf 64}$\\
$4$ & $2\times {\bf 1},\;\, 4\times {\bf 8},\;\,{\bf 10},\;\,\overline{\bf 10},\;\, 3\times {\bf 27},\;\, {\bf 35},\;\, \overline{\bf 35},\;\, {\bf 64},\;\, {\bf 125}$ \\
\hline
\end{tabular}
\caption{The multiplicities of $su(3)$ representations at the first five levels in \eqref{eq:A1D4-index}.}
\label{table:A1D4-multiplicities}
\end{center}
\end{table}
Analogously to the case of the $(A_1, A_3)$ theory, here we would like to compare the normalized torus partition function (or character) of the vacuum module of $\widehat{su}(3)_{-{3\over2}}$, $\CH_0^{[-{3\over2},0,0]}$, with our conjecture for the Schur index. Note that the vacuum module has highest weight, $\lambda=[-\frac{3}{2},0,0]$. In particular, we should have
\begin{align}
\mathcal{I}_{(A_1,D_4)}(q;\vec{z}) = {\rm Tr}_{\CH_0^{[-{3\over2},0,0]}}(-1)^F q^{L_0}\prod_{i=1,2}(z_i)^{f_i}\equiv e^{-\lambda}\text{ch}\, L(\lambda)~.
\label{eq:conjecture}
\end{align}
The fugacities are identified as $q = e^{-\delta},\; z_1 = e^{-(\frac{2}{3}\alpha_1+\frac{1}{3}\alpha_2)}$, and $z_2=e^{-(\frac{1}{3}\alpha_1+\frac{2}{3}\alpha_2)}$, where the notation is summerized in appendix \ref{app:character}. The character formula of \rcite{Kac-Wakimoto} then tells us that 
\begin{align}
 e^{-\lambda}\text{ch}\, L(\lambda) =  1 \,+\, \chi_{\bf 8}^{su(3)}(\vec{z})\,q \,+\, \big[1 + \chi_{\bf 8}^{su(3)}(\vec{z}) + \chi_{\bf 27}^{su(3)}(\vec{z})\big]q^2  \,+\, \cdots,
\label{eq:A1D4-index}
\end{align}
where $\chi_{\bf n}^{su(3)}(\vec{z})$ are characters of $su(3)$.\footnote{To be explicit, for an $su(3)$ representation with the Dynkin label $(\lambda_1,\lambda_2)$, the character is written as $\chi_{(\lambda_1,\lambda_2)}^{su(3)}(\vec{z})= \det(w_i^{\ell_j+3-j})/\det(w_i^{3-j})$ where $w_1=z_1,\,w_2=1/z_2,\,w_3=z_2/z_1$, and $\ell_1=\lambda_1+\lambda_2,\; \ell_2=\lambda_2,\;\ell_3=0$.} Note that the $SU(3)$ flavor symmetry of the theory is manifest in this expression. The multiplicities of the $su(3)$ representations up to $\mathcal{O}(q^4)$ are summarized in Table \ref{table:A1D4-multiplicities}.

Now, our conjecture \eqref{eq:A1Dn-index} implies that the same Schur index is written as
\begin{align}
\mathcal{I}_{(A_1,D_4)}(q;x,y) = \sum_{R} \tilde{f}^{(2)}_{R}(q;x)f_R(q;y)~.
\label{eq:conjecture2}
\end{align}
Here $x$ and $y$ are the fugacities for mutually commuting $U(1)$ and $SU(2)$ subgroups of the $SU(3)$ flavor symmetry, which are related to the $z_i$ by the map $x=(z_2)^{-\frac{3}{2}}$ and $y=z_1(z_2)^{-\frac{1}{2}}$.
We have checked that \eqref{eq:conjecture2} and \eqref{eq:A1D4-index} match up to high perturbative order in $q$ under this identification of the fugacities.
This result is a highly non-trivial consistency check of our conjecture.
Crucially, this result implies that the manifest $U(1)\times SU(2)$ flavor symmetry in \eqref{eq:conjecture2} is enhanced to $SU(3)$.

\subsubsection{Comments on the chiral algebras of general AD theories}
\label{subsubsec:general-AD}

Let us briefly comment on some aspects of chiral algebras for more general AD theories (including the $(A_1, A_{2n-3})$ and $(A_1, D_{2n})$ SCFTs). As reviewed around \eqref{torusIndex}, once the two-dimensional chiral algebra, $\chi_{\mathcal{T}}$, is identified, we can in principle compute the Schur index of the corresponding four-dimensional theory, $\mathcal{T}$ (although, as we saw in the cases of the $(A_1, A_3)$ and $(A_1, D_4)$ SCFTs, our closed-form expressions for the indices remain highly non-trivial even once we know the corresponding chiral algebras). While we do not have a complete algorithm to find $\chi_{\CT}$ for a generic AD theory, we will argue that $\chi_{\CT}$ is highly constrained when $\mathcal{T}$ is an AD theory.

To that end, we begin by noting that any four-dimensional generator in a certain subclass of Schur operators called the Hall-Littlewood (anti)chiral ring is guaranteed to be a chiral algebra generator (by a simple argument in representation theory\rcite{Beem:2013sza}).\foot{By chiral algebra generators, we mean $sl(2)$ primaries who, along with their ($sl(2)$) descendants and the resulting normal-ordered products, span $\chi_{\CT}$.} In what follows, an even more special (closed) subclass of the Hall-Littlewood (anti)chiral ring, the Higgs branch (anti)chiral ring, will be particularly important. These operators are highest weight $\CN=2$ primaries, $\CO^{1\cdots 1}$, (the indices are $SU(2)_R$ indices set to the highest weight; the lowest weight states are in the conjugate ring) transforming in short multiplets of type $\hat\CB_R$ in the language of \rcite{Dolan:2002zh}.\foot{More precisely, they are operators satisfying $\left[\CQ^1_{\alpha},\CO^{1\cdots1}\right]=\left[\tilde \CQ_{2\dot\alpha},\CO^{1\cdots1}\right]=0$.} 
For example, given a $G_F$ flavor symmetry in four dimensions, there is always a corresponding conserved current multiplet of type $\hat B_1$ whose Schur operator is the holomorphic moment map, $\CO^{11}=M^{11}$ (the conserved current is a descendant). The two-dimensional image of this moment map (or, more precisely, of the element in a certain twisted cohomology corresponding to this moment map \rcite{Beem:2013sza}) is the current that generates the AKM algebra, $M(z, \bar z)$. In particular, the corresponding self-OPEs in four and two dimensions are
\eqn{
M_A^{ij}(x)M_B^{kl}(0)\sim -{3k_{4d}\over48\pi^4}{\epsilon^{k(i}\epsilon^{j)l}\over x^4}\delta_{AB}-{\sqrt{2}\over4\pi^2}{f_{ABC}\over x^2}M_C^{(i(k}(0)\epsilon^{l)j)}+\cdots~,
}[mmOPE4d]
and
\eqn{
M_A(z, \bar z)M_B(0,0)\sim~{3k_{2d}\over24\pi^4}{1\over z^2}\delta_{AB}+{\sqrt{2}\over4\pi^2}{if_{ABC}\over z}M_C(0,0)+{\sqrt{2}\over4\pi^2}{\bar z\cdot f_{ABC}\over z}M_{C}^{21}(0)+\cdots~,
}[mmOPE2d]
where the last term in \eqref{mmOPE2d} and, in fact, all $\bar z$-dependent terms, are $Q$-exact with respect to a nilpotent supercharge specified in \rcite{Beem:2013sza}. In these two equations, the subscripts label the generator of $G_F$, and the superscripts are $SU(2)_R$ indices.

For our purposes, there are three crucial observations regarding \eqref{mmOPE2d}. First, a rather basic fact is that $M_A$ is {\it not} an AKM primary. Instead, it is an AKM descendant of the identity. As a result, the AKM currents are in the vacuum module of the AKM algebra (we used this fact in the previous two subsections). On the other hand, this operator {\it is} a generator of the chiral algebra (and, in fact, is a Virasoro primary). The second observation is that the OPE in \eqref{mmOPE2d} is single-valued. Finally, we note that the meromorphic contributions to \eqref{mmOPE2d} only involve two-dimensional cousins of Schur operators. In fact, these last two observations are general \rcite{Beem:2013sza}. Indeed, for any two operators, $\CO_{1,2}$, in $\chi_{\CT}$, we have
\eqn{
\CO_1(z,\bar z)\CO_2(0,0)=\sum_{k_{\rm Schur}}{\lambda_{12k}\over z^{h_1+h_2-h_k}}\CO_k(0)+\{Q,\cdots]~.
}[GenOPE]
Single-valuedness of the non-$Q$-exact part of the OPE then follows from superconformal representation theory \rcite{Beem:2013sza}.

As discussed above, in the cases of the $(A_1,A_3)$ and $(A_1,D_4)$ theories, the AKM generators, $M_A$, are expected to generate $\chi_{\mathcal{T}}$. On the other hand, in the cases of $(A_1,A_{2n-3})$ for $n>3$ and $(A_1,D_{2n})$ for $n>2$, there are Hall-Littlewood operators which are not generated by the flavor moment maps. Such operators map to chiral operators which are not generated by $M_A$ in two dimensions. Therefore $\chi_{\mathcal{T}}$ now contains more than the AKM vacuum module. In particular, it must at least contain modules that include two-dimensional cousins of certain four-dimensional baryons (we will introduce these operators in section 5).

Now, we would like to argue that the AKM modules are highly constrained for generic AD theories with non-abelian flavor symmetry. To understand this statement recall that for a generic AD theory with some non-abelian flavor symmetry, we expect to have an AKM sub-theory with a fractional level. However, for generic fractional levels, these AKM theories are believed to admit logarithmic representations (LRs) (see, e.g., the recent review \rcite{Gaberdiel:2001tr, Ridout:2015fga} and references therein; note that this discussion does not preclude LRs in theories with negative integer levels). For example, it is known that $\widehat{su}(2)_{-{4\over3}}$ admits such representations \rcite{Gaberdiel:2001ny}. Moreover, fusion rules suggest that LRs appear once we add too many non-logarithmic representations. Since the LRs have logarithms in their correlation functions, these results are somewhat in tension with the fact, discussed around \eqref{GenOPE}, that the meromorhic OPE should be single-valued. Of course, there is no contradiction with our proposal in the case of the $(A_1, A_3)$ theory, since we are only claiming that the chiral algebra consists of the vacuum module of $\widehat{su}(2)_{-{4\over3}}$ (this module does not give rise to logarithmic correlation functions and is closed under fusion).

But this tension points to a more general heuristic picture. If an AD theory has some non-abelian flavor symmetry, $G_F$, then the sector of operators charged under $G_F$ will likely be relatively simple in the following sense: these four-dimensional operators should correspond to two-dimensional operators transforming in a relatively small number of irreducible AKM modules. Otherwise, the fusion rules of the chiral algebra may lead to modules that give rise to correlation functions with logarithms. As a result, this discussion points to another way in which AD theories behave more simply than one might expect. Moreover, the behavior discussed here is clearly complementary to our characterization of the relative simplicity of AD theories discussed around \eqref{aVsRank}, which was based on properties of the Coulomb branch.

\subsection{Consistency with S-duality}

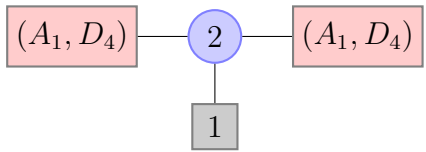
\begin{figure}
\begin{center}
\vskip .5cm
\begin{tikzpicture}[place/.style={circle,draw=blue!50,fill=blue!20,thick,inner sep=0pt,minimum size=7mm},transition/.style={rectangle,draw=black!50,fill=black!20,thick,inner sep=0pt,minimum size=6mm},transition2/.style={rectangle,draw=black!50,fill=red!20,thick,inner sep=0pt,minimum size=8mm},auto]

\node[transition2] (1) at (1.1,0) {\;$(A_1,D_4)$\;};
\node[place] (2) at (3,0) [shape=circle] {$2$} edge [-] node[auto]{} (1);
\node[transition2] (3) at (4.9,0)  {\;$(A_1,D_4)$\;} edge [-] node[auto]{} (2);
\node[transition] (9) at (3,-1.2) {$1$} edge[-] (2);
\end{tikzpicture}
\caption{The $(A_3,A_3)$ theory is equivalent to the above quiver gauge theory in which a diagonal $SU(2)$ flavor symmetry of the two $(A_1,D_4)$ theories are gauged by an $SU(2)$ vector multiplet coupled to a fundamental hypermultiplet.}
\label{fig:quiver}
\end{center}
\end{figure}

In this sub-section, we perform another consistency check of our conjecture \eqref{eq:A1Dn-index} for the $(A_1,D_4)$ index using an $S$-duality discovered in \rcite{Buican:2014hfa}. To that end, we start with the theory of two $(A_1,D_4)$ SCFTs and a doublet of free hypermultiplets. The resulting flavor symmetry is $SU(3) \times SU(3)\times Sp(2)$ of which we gauge a diagonal $SU(2)$ subgroup. The matter content turns out to be precisely right so that this gauging is exactly marginal. The resulting conformal manifold of SCFTs has $U(1)^3$ flavor symmetry and is described by the quiver diagram in figure \ref{fig:quiver}. This theory is known to be identical to the so-called $(A_3,A_3)$ theory \rcite{Cecotti:2010fi}.

Let us now consider the Schur index of the $(A_3,A_3)$ theory.
Since the superconformal index is invariant under exactly marginal deformations, it can be evaluated in a weak gauge coupling limit (note, however, that the full theory still consists of strongly coupled sub-sectors) \rcite{Romelsberger:2005eg, Kinney:2005ej} using our conjecture in \eqref{eq:conjecture2} as input
\begin{align}
\mathcal{I}_{(A_3,A_3)} (q;x,y,z) = \oint& \frac{dw}{2\pi i w}\Delta(w)\;\mathcal{I}^{SU(2)}_{\text{vect}}(q;w)\;\mathcal{I}^{SU(2)}_{\text{fund}}(q;x,w)\; \mathcal{I}_{(A_1,D_4)}(q;y,w)\;\mathcal{I}_{(A_1,D_4)}(q;z,w)~,
\label{eq:A3A3-integral}
\end{align}
where the measure factor is given by $\Delta(w)=\frac{1}{2}(1-w^2)(1-w^{-2})$, and
\begin{align}
&\mathcal{I}_{\text{vect}}^{SU(2)}(q;w) = P.E.\left[-\frac{2q}{1-q}\chi_{\bf 3}^{su(2)}(w)\right]~,
\nonumber \\[1mm]
&\mathcal{I}_{\text{fund}}^{SU(2)}(q;x,w) = P.E.\left[\frac{\sqrt{q}}{1-q}(x+x^{-1})\chi_{\bf 2}^{su(2)}(w)\right]~,
\end{align}
are the contributions from the vector and the doublet of hypermultiplets, respectively. The fugacities $x,y,z$ are associated with the $U(1)^3$ flavor symmetry of the $(A_3,A_3)$ theory (recall that the parameters $x$ and $w$ of $\mathcal{I}_{(A_1,D_4)}(q;x,w)$ are, respectively, the flavor fugacities for the commuting $U(1)$ and $SU(2)$ flavor subgroups of the $(A_1,D_4)$ theory).

One of the results of \rcite{Buican:2014hfa} is the discovery of an $S$-duality for the $(A_3,A_3)$ theory.\footnote{The $(A_k,A_k)$ theory is called the ``$I_{k+1,k+1}$ theory'' in \rcite{Buican:2014hfa}.} This $S$-duality exchanges the strong and weak gauge coupling limits together with permuting the parameters of the theory. Let $m_x$ be the mass for the doublet of hypermultiplets, and $m_y,m_z$ be those for the $(A_1,D_4)$ sectors. These masses are mixed as $m_x \to \frac{1}{2}(m_y-m_z),\; m_y \to m_x+\frac{1}{2}(m_y+m_z),\; m_z \to -m_x+\frac{1}{2}(m_y+m_z)$ under the $S$-duality transformation.\footnote{The physical masses $m_1,m_2$, and $m_3$ discussed in \rcite{Buican:2014hfa} are related to these masses by the mapping $m_1 = 2m_x,\, m_2=m_y+m_z$, and $m_3=m_y-m_z$ (modulo permutations).} This action corresponds to permuting the generators of the $U(1)^3$ flavor symmetry group. In terms of the fugacities $x,y$ and $z$, this permutation is expressed as
\begin{align}
x \to \sqrt{y/z}~,\ \ \ y\to x\sqrt{yz}~,\ \ \  z\to\frac{\sqrt{yz}}{x}~.
\label{eq:transformation}
\end{align}
Since the theory is invariant under $S$-duality, the superconformal index should also be invariant. This observation implies the following relation:
\begin{align}
\mathcal{I}_{(A_3,A_3)}(q;x,y,z) \;=\; \mathcal{I}_{(A_3,A_3)}\Big(q;\sqrt{y/z},\, x\sqrt{y z},\, \frac{\sqrt{y z}}{x}\Big)~.
\end{align}
We have checked that \eqref{eq:A3A3-integral} combined with our conjecture, \eqref{eq:conjecture2}, correctly satisfies this identity up to a high perturbative order in $q$. This invariance is strong evidence for our conjecture. Note that, combined with our discussion in the previous sub-section, this result also supports the chiral algebra conjecture $\chi_{(A_1, D_4)}=\widehat{su(3)}_{-{3\over2}}$.

\section{General consistency}

In this section, we will perform various consistency checks available for general $(A_1,A_{2n-3})$ and $(A_1,D_{2n})$ theories.

\subsection{Flavor symmetry and Higgs branch relations}

While the full spectrum of Schur operators is not known for general $(A_1,A_{2n-3})$ and $(A_1,D_{2n})$ theories, there are several such operators that are universal. The most obvious example of a universal Schur operator is the highest-weight component of the $SU(2)_R$ current in the stress tensor multiplet. In addition, if the theory has $G_F$ flavor symmetry, then, as we discussed in section 4.2, there exist flavor moment map Schur operators which transform in the adjoint representation of $G_F$. Since these operators have $E=2R=2$, their contribution to the index is
\begin{align}
q\,\chi_{\text{adj}}^{G_F}(\vec{x})~,
\end{align}
where $\vec{x}$ is the flavor fugacity. In fact, this is the only representation that can contribute linearly in $q$ (this conclusion follows from studying the full set of short multiplets in \rcite{Dolan:2002zh}). As we mentioned in section 4.2, the flavor moment maps are examples of Higgs branch operators, i.e., operators whose vevs parameterize the Higgs branch.

More generally, the Higgs branch operators are all Schur operators with $E=2R$. As a result, they first contribute to the index at $\mathcal{O}(q^R)$.  Below, we check that our conjectures \eqref{eq:A1An-indexii} and \eqref{eq:A1Dn-indexii} correctly reproduce these contributions.

\subsubsection{The $(A_1,A_{2n-3})$ theories}

\begin{table}
\begin{center}
\begin{tabular}{c|c|c|c}
 & $E$ & $R$ & $U(1)$ charge\\ \hline
$M$ & $2$ & $1$ & $0$ \\[1mm]
$N^\pm$ & $n-1$ & $\frac{n-1}{2}$ & $\pm 1$ \\
\end{tabular}
\caption{The dimensions, the $SU(2)_R$ charges, and the flavor $U(1)$ charges of the Higgs branch operators of the $(A_1,A_{2n-3})$ SCFT.}
\label{table:A1An-Higgs}
\end{center}
\end{table}
Let us first study the $(A_1,A_{2n-3})$ theories. We assume $n\geq 3$ so that the theory is interacting. As discussed above, the flavor symmetry of the theory is $U(1)$ for $n\geq 4$ and $SU(2)$ for $n=3$. The Higgs branch of the theory is known to be the orbifold ${\bf C}^2/{\bf Z}_{n-1}$, and is therefore parameterized by the vevs of three Higgs branch operators $N^+$, $N^-$, and $M$ constrained by \rcite{Argyres:2012fu}
\begin{align}
N^+N^- = (-1)^{[\frac{n-1}{2}]}M^{n-1}~.
\label{eq:A1An-Higgs}
\end{align}
The dimensions, $SU(2)_R$ charges, and $U(1)$ flavor charges of these operators are summarized in Table \ref{table:A1An-Higgs}. 
The flavor moment map operators are $M$ for $n>3$ and $(M,N^\pm)$ for $n=3$.

Let us check that our conjecture for the Schur index is consistent with these features of the theory. First of all, the $\mathcal{O}(q)$ contribution to our index is given by
\begin{align}
\mathcal{I}_{(A_1,A_{2n-3})}(q;x)\Big|_{\mathcal{O}(q)} = \left\{
\begin{array}{l}
q\qquad\qquad\quad \text{for}\quad n\geq 4\\
q\,\chi_{\bf 3}^{su(2)}(\sqrt{x})\qquad \text{for}\quad n=3\\
\end{array}
\right.,
\end{align}
which is the correct contribution from the $U(1)$ or $SU(2)$ flavor moment map operators. For $n\geq 4$, the $N^\pm$ are not the flavor moment maps and contribute to the index at $\mathcal{O}(q^{\frac{n-1}{2}})$. It is easy to see that our $\mathcal{I}_{(A_1,A_{2n-3})}(q;x)$ always contains the corresponding terms $q^{\frac{n-1}{2}}(x+x^{-1})$.

Our index is also consistent with the Higgs branch relation \eqref{eq:A1An-Higgs}. To see this, let us rewrite our Schur index as
\begin{align}
\mathcal{I}_{(A_1,A_{2n-3})}(q;x) =& P.E.\left[i_T(q)\right]P.E.\left[i_{M}(q)\right]P.E.\left[i_{N}(q;x)\right]\left(1  - \sum_{k>0} r_k(x)q^k\right),
\end{align}
where $i_T(q) = q^2/(1-q),\; i_M(q) = q/(1-q),\; i_{N}(q;x) = q^{\frac{n-1}{2}}(x+x^{-1})/(1-q)$ are the indices of the stress tensor multiplet and the Higgs branch operators (along with their derivatives). The functions $r_{k}(x)$ contain information about operator relations as well as the index contributions from Schur operators which are not generated by $T,M,N^\pm$. Since the Higgs branch relation \eqref{eq:A1An-Higgs} has $E-R=n-1$, its effect should be encoded in $r_{n-1}(x)$. Indeed, our conjecture implies
\begin{align}
r_{n-1}(x) = 1~,
\end{align}
which is consistent with the Higgs branch relations \eqref{eq:A1An-Higgs}: it eliminates one of the two separate contributions from $N^+N^-$ and $M^{n-1}$. Note here that, in the case $n=3$, this Higgs branch relation becomes the Joseph ideal constraint \rcite{Gaiotto:2008nz} associated with the enhanced $SU(2)$ flavor symmetry.\foot{Before considering the $(A_1, D_{2n})$ theories, we should mention a word of caution. While we have seen that our formula for the index naturally includes the expected Higgs branch contributions at $\CO(q^{n-1\over2})$ and also includes the expected constraints \eqref{eq:A1An-Higgs}, we should note that we have not yet given definitive proof that this is the case (although our statement for the $\CO(q^{n-1\over2})$ contributions can be regarded as rigorously proven when $n=3$, since this statement follows from the general discussion in \rcite{Dolan:2002zh}). Indeed, the index is ambiguous in the sense that different contributions to it can sometimes cancel.  As a result, while we have given the most obvious interpretation consistent with our conjectured form of the index and the known facts about the Higgs branch of the $(A_1, A_{2n-3})$ theory, to prove this interpretation is correct we need to learn more about the spectrum of the theory (for example, by constructing the relevant chiral algebra). We will return to these questions in an upcoming paper \rcite{toappear}.\label{Opsubtlety}}

\subsubsection{The $(A_1,D_{2n})$ theories}

\begin{table}
\begin{center}
\begin{tabular}{c|c|c|c}
& $E$ & $R$ & rep. of $SU(2)\times U(1)$ \\\hline
$M_i{}^j$ & $2$ & $1$ & $({\bf 2}\otimes {\bf 2})_0 = {\bf 3}_0 \oplus {\bf 1}_0$ \\[.5mm]
$L_i$ & $n$ & $\frac{n}{2}$ & ${\bf 2}_{1}$ \\
$\tilde{L}^i$ & $n$ & $\frac{n}{2}$ & ${\bf 2}_{-1}$ \\
\end{tabular}
\caption{The charges of the Higgs branch operators of the $(A_1,D_{2n})$ theories. Here $i,j=1,2$ are spinor indices associated with the flavor $SU(2)$.}
\label{table:A1Dn-Higgs}
\end{center}
\end{table}

We now turn to the $(A_1,D_{2n})$ theories. We assume $n\geq 2$ so that the theory is interacting. As described above, the flavor symmetry of the theory is $SU(2)\times U(1)$ for $n\geq 3$ and $SU(3)$ for $n=2$. Moreover, the theory has a four quaternionic dimensional Higgs branch parameterized by the vevs of the Higgs branch operators $L_i,\,\tilde{L}^i$, and $M_i{}^j$ for $i,j=1,2$. Their dimensions, $R$-charges, and flavor charges are summarized in Table \ref{table:A1Dn-Higgs}. The relations among these operators can be read off from the discussion of the $S^1$ reduction in the next sub-section
\begin{align}
\epsilon^{ik}\epsilon_{j\ell}M_i{}^j M_k{}^\ell =0~, \ \ \ \epsilon_{jk}M_i{}^j\tilde{L}^k =0~, \ \ \  \epsilon^{ik}M_i{}^jL_k =0,\quad L_i\tilde{L}^j = (-1)^{[\frac{n}{2}]}M_i{}^j(M_k{}^k)^{n-1}~.
\label{eq:A1Dn-Higgs}
\end{align}
Note that the above four relations transform in the representations ${\bf 1}_0,\, {\bf 2}_{-1},\, {\bf 2}_{1}$, and ${\bf 3}_0\oplus{\bf 1}_0$ of the $SU(2)\times U(1)$ flavor symmetry. The relations also have $E-R= 2,\, \frac{n}{2}+1,\, \frac{n}{2}+1$, and $n$, respectively. The flavor moment maps are $M_i{}^j$ for $n\geq 3$ and $M_i{}^j,\,L_i,\,\tilde{L}^i$ for $n=2$.

Let us now examine our conjecture \eqref{eq:A1Dn-indexii} for the Schur index. The $\mathcal{O}(q)$ terms in the index are written as
\begin{align}
\mathcal{I}_{(A_1,D_{2n})}(q;x,y)\Big|_{\mathcal{O}(q)} = \left\{
\begin{array}{l}
q\,(\chi_{\bf 3}^{su(2)}(y)+1) \qquad \qquad \text{for}\quad n\geq 3\\[1mm]
q\,\chi_{\bf 8}^{su(3)}(x^{-\frac{1}{3}}y, x^{-\frac{2}{3}})\qquad \qquad \text{for}\quad n=2\\
\end{array}
\right.,
\end{align}
which are the correct contributions from the flavor moment map operators. For $n\geq 3$, the $L_i,\,\tilde{L}^i$ are not flavor moment maps. Instead, their contribution is $\mathcal{O}(q^{\frac{n}{2}})$. Indeed, one can check that $\mathcal{I}_{(A_1,D_{2n})}(q;x,y)$ always contains the corresponding terms, $q^{\frac{n}{2}}\chi_{\bf 2}^{su(2)}(y)(x+x^{-1})$.

To see that our conjecture is consistent with the Higgs branch relations \eqref{eq:A1Dn-Higgs},  let us rewrite our index as
\begin{align}
\mathcal{I}_{(A_1,D_{2n})}(q;x,y) =& P.E.\left[i_T(q)\right]P.E.\left[i_M(q)\right]P.E.[i_{L}(q;x,y)] \left(1 - \sum_{k>0}r_k(x,y)q^k\right)~,
\label{eq:A1Dn-relation}
\end{align}
where $i_T(q) = q^2/(1-q)$, $i_M(q) = q(\chi_{\bf 3}^{su(2)}(b) + 1)/(1-q)$, and $i_{L}(q;x,y) = q^{\frac{n}{2}}\chi_{\bf 2}^{su(2)}(y)(x+x^{-1})/(1-q)$ are the contributions to the index from the stress tensor multiplet and the Higgs branch operators. The operator relations are encoded in $r_k(x,y)$. In the case $n\geq 3$, we have
\begin{align}
r_2(x,y)  = 1~,\ \ \  r_{\frac{n}{2}+1}(x,y) = \chi_{\bf 2}^{su(2)}(y)(x+x^{-1})~,
\end{align}
which are consistent with the first three of the Higgs branch relations \eqref{eq:A1Dn-Higgs}. The last Higgs branch relation affects $r_n(x,y)$, whose expression is generically not simple but always contains the terms $\chi_{\bf 3}^{su(2)}(y) + 1$. This contribution is consistent with the last Higgs branch relation in \eqref{eq:A1Dn-Higgs}.
In the case of $n=2$, all of the relations in \eqref{eq:A1Dn-Higgs} contribute to $r_2(x,y) = 1 + \chi_{\bf 2}^{su(2)}(y)(x+x^{-1}) + \chi_{\bf 3}^{su(2)}(y) + 1= \chi_{\bf 8}^{su(3)}(x^{-\frac{1}{3}}y,x^{-\frac{2}{3}}) + 1$. This case corresponds to the Joseph ideal constraints on the Higgs branch operators of the $(A_1,D_4)$ theory \rcite{Gaiotto:2008nz}.\footnote{Note that the somewhat overly cautious discussion in footnote \ref{Opsubtlety} applies in the case of the $(A_1, D_{2n})$ theory as well.}

\subsection{The RG flow}
\label{subsec:Higgsing}
In this sub-section, we describe three types of RG flows that interpolate between the above theories. These flows provide additional consistency checks of our conjectures via the mechanism described in  \rcite{Gaiotto:2012xa}. In particular, our RG flows are initiated by turning on vevs for certain Higgs branch operators. In the language of the index, turning on these vevs corresponds to taking a certain singular limit of fugacities. The index of the IR theory can then be read off from the resulting residues \rcite{Gaiotto:2012xa}.

As a check of these computations, we give an alternate description of the RG flow at the level of the $S^1$ reduction. While it is true that this three-dimensional description of the flow may not correspond to a unique RG flow in four dimensions (for example, given one of the AD theories we study here, there is always a certain abelian gauge theory of the same rank in four dimensions that reduces to the same theory in three dimensions\foot{Although,  in an upcoming work \rcite{Buican:2015hsa}, we will see that there is a subtle difference at the level of the $S^1$ reductions of the corresponding indices.}), our flows in four dimensions are special: they do not break the $U(1)_R\subset U(1)_R\times SU(2)_R$ $R$ symmetry of the UV theory, since the Higgs branch operators are uncharged under $U(1)_R$. As a result, we expect the Coulomb branch spectrum of the IR theory to be a subset of the Coulomb branch spectrum of the UV theory. Therefore, when the IR theory has a non-trivial Coulomb branch in three dimensions, we can determine its four-dimensional parent based on the Coulomb branch spectrum (i.e., we can discriminate between AD theories and abelian gauge theories of the same rank in four dimensions by the scaling dimensions of the chiral operators).
On the other hand, when the IR theory has no Coulomb branch in three dimensions, this situation corresponds to a theory of rank zero in four dimensions, and the only such theories are believed to be theories of free hypermultiplets. Therefore, the consistency of this three-dimensional picture with the index computation is strong evidence for our conjecture.

\subsubsection{The $(A_1, A_{2n-3})\to (A_1, A_1)$ RG flow}

To understand the above statements further, let us begin by studying the RG flow
\eqn{
(A_1, A_{2n-3})\to (A_1, A_1)~.
}[A1A2n3A1A1]
We claim that we can initiate \eqref{A1A2n3A1A1} by, for example, turning on $\langle N^+\rangle\ne0$ while keeping $\langle N^-\rangle=\langle M\rangle=0$.\foot{It is also possible to turn on $\langle N^+\rangle$ with $\langle N^-\rangle=\langle M\rangle=0$ instead. Finally, we can take $\langle N^{\pm}\rangle,\langle M\rangle\ne0$ as long as we satisfy \eqref{eq:A1An-Higgs}.} As discussed above, one way to see this flow explicitly is to consider the $S^1$ reduction of the $(A_1, A_{2n-3})$ theory. The endpoint of the resulting RG flow is an interacting three-dimensional $\CN=4$ SCFT. This latter theory can alternatively be reached by starting with an abelian three-dimensional $\CN=4$ gauge theory whose matter content is summarized in Table \ref{AAside} (note, however, that in the flow from the AD theory in four dimensions, we never land on the free three-dimensional theory).\foot{The authors of \rcite{Argyres:2012fu} found a similar effective theory written in terms of four-dimensional gauge and matter fields at a point parametrically close (in the Coulomb branch) to the AD point. Much of our three-dimensional discussion can be rephrased in terms of this effective theory.}
\begin{table}
\begin{center}
     \begin{tabular}{| c | c | c | c | c | c |}
\hline   & $U(1)_1$ & $U(1)_2$ & $\cdots$ & $U(1)_{n-3}$& $U(1)_{n-2}$\cr\hline\hline
        $q_1$ & $1$ & $0$& $\cdots$ & 0 & 0\cr\hline
        $q_2$ & $1$ & $1$& $\cdots$ & 0 & 0\cr\hline
        $\vdots$ & $\vdots$ & $\vdots$& $\ddots$ & $\vdots$ & \vdots\cr\hline
        $q_{n-2}$ & $0$ & $0$& $\cdots$ & 1 & 1\cr\hline
        $q_{n-1}$ & $0$ & $0$& $\cdots$ & 0 & 1\cr\hline
      \end{tabular}
\caption{The matter fields (and charges) of the three-dimensional gauge theory that flows to the $S^1$ reduction of the $(A_1,A_{2n-3})$ theory.}
\label{AAside}
\end{center}
\end{table}
This gauge theory has a superpotential\foot{According to the discussion in \rcite{Xie:2012hs, Boalch:XXXXxx, Boalch:YYYYyy}, the IR fixed point of $\CN=4$ SQED with $N_f=n-1$ gives the mirror of the reduction of the $(A_1, A_{2n-3})$ theory, and so we can derive the above theory by applying mirror symmetry again.}
\eqn{
W=\sum_{a=1}^{n-2}\Phi_a(q_{a}\tilde q^{a}+q_{a+1}\tilde q^{a+1})~.
}[A1A2m3superpot]
As a result, the Higgs branch of the theory described in Table \ref{AAside} captures the Higgs branch of the $(A_1, A_{2n-3})$ SCFT, and we can map the Higgs branch operators between the theories as follows\foot{We assume that $n$ is even in \eqref{HiggsBrMapA1A2n3}. In the case of odd $n$, we exchange $q_{n-1}\leftrightarrow\tilde q^{n-1}$.}
\eqn{
N^+\leftrightarrow q_1\tilde q^2 q_3\cdots q_{n-1}~, \ \ \ N^-\leftrightarrow \tilde q^1 q_2\tilde q^3\cdots \tilde q^{n-1}~, \ \ \ M\leftrightarrow q_1\tilde q^1~.
}[HiggsBrMapA1A2n3]
Note that the $U(1)$ flavor symmetry of the $(A_1, A_{2n-3})$ theory manifests itself as the flavor symmetry under which $q_1$ has charge $+1$, $\tilde q^1$ has charge $-1$, and the rest of the fields are neutral (in the case of $n=3$, the $q_i$ transform as a doublet under the enhanced $SU(2)$ symmetry).

Clearly, if we turn on $\langle N^+\rangle\ne0$, then we completely break the $U(1)^{n-2}$ gauge symmetry. The corresponding gauge bosons become massive and eat all but one of the hypermultiplet degrees of freedom. This remaining hypermultiplet is the axion-dilaton multiplet and comprises (the $S^1$ reduction of) the $(A_1, A_1)$ theory. In terms of the class $\CS$ description, we can think of this RG flow as Higgsing the irregular puncture of the $(A_1, A_{2n-3})$ theory. It is then reasonable for us to be left over with the axion-dilaton and an empty theory in the IR.

Next, let us describe this RG flow from the perspective of the index. To that end, recall the prescription given in \rcite{Gaiotto:2012xa} for relating the UV and IR indices of the RG flow induced by turning on the vev of a Higgs branch operator, $\CO$, of charge $f_{k, \CO}$ under some $U(1)_k$ flavor symmetry (this symmetry may correspond to a Cartan of a non-abelian symmetry) and $SU(2)_R$ weight $R_{\CO}$
\eqn{
\CI_{\rm vect}^{-1}\cdot\CI_{IR}=-f_{i,\CO}\cdot {\rm Res}_{x_i=q^{-{R_{\CO}\over f_{\CO}}}\prod_{j\ne i}x_j^{-{f_{j,\CO}\over f_{i,\CO}}}}\left({1\over x_i}\CI_{UV}\right)~.
}[GRR]
In \eqref{GRR}, $x_k$ is a fugacity for the flavor $U(1)_k$, $\mathcal{I}_{vect}$ is the index of a free $U(1)$ vector multiplet, and $\mathcal{I}_{UV}$ is the index of the UV theory. Note that $\CI_{IR}$ is the index of all the IR degrees of freedom {\it except} the decoupled axion-dilaton (the axion-dilaton multiplet contribution is $\CI_{\rm vect}^{-1}$, since it would disappear from the spectrum upon gauging the $U(1)$). For the flow \eqref{A1A2n3A1A1}, the IR theory is trivial (since the only degrees of freedom are the axion-dilaton), and so $\CI_{IR}=1$. Moreover, $f_{N^+}=1$ while $R_{N^+}={n-1\over2}$. As a result, since the UV index is $\mathcal{I}_{(A_1,A_{2n-3})}$, we need to compute the residue of $x^{-1}\CI_{(A_1, A_{2n-3})}$ at $x=q^{-{n-1\over2}}$. This calculation is straightforward using the following re-writing of our conjecture \eqref{eq:A1An-index}
\begin{eqnarray}\label{A1A2n3index}
\CI_{(A_1, A_{2n-3})}(q;x)&=&{1-q\over(q;q)_{\infty}^2}\sum_{k=0}^{\infty}\Big({1\over q^{1\over2}-q^{-{1\over2}}}\sum_{s=\pm1}\left[{q^{n(2k+1)^2-2nk^2+2k+2\over2}x^s\over1-q^{(k+{1\over2})n+{1\over2}}x^s}-{q^{n(2k+1)^2-2nk^2-2k-2\over2}x^s\over1-q^{(k+{1\over2})n-{1\over2}}x^s}\right]\nonumber\\ &+& q^{nk(k+1)}{q^{k+{1\over2}}-q^{-k-{1\over2}}\over q^{1\over2}-q^{-{1\over2}}}\Big)~.
\end{eqnarray}
Indeed, we immediately see that
\eqn{
\CI_{\rm vect}^{-1}(q)\CI_{\emptyset}=\CI_{\rm vect}^{-1}(q)=-{\rm Res}_{x=q^{-{n-1\over2}}}\left({1\over x}\CI_{(A_1, A_{2n-3})}(q;x)\right)~,
}[RGtoA1A1]
which is in perfect agreement with \eqref{GRR}.
This is a highly non-trivial check of our conjecture.

\subsubsection{The $(A_1, D_{2n})\rightarrow(A_1, A_{2n-3})\oplus(A_1, A_1)$ RG flow}

Next, we argue that, in addition to the RG flow in \eqref{A1A2n3A1A1}, we also have
\eqn{
(A_1, D_{2n})\rightarrow(A_1, A_{2n-3})\oplus(A_1, A_1)~.
}[A1D2ntoA1A2n3]
We claim that we can initiate this flow by, for example, turning on $\langle M_1^{\ 2}\rangle\ne0$ (and keeping the vevs of all the other operators in Table \ref{table:A1Dn-Higgs} vanishing).\foot{We can consider many other possible Higgsings that result in the same flow. Indeed, we need only satisfy \eqref{eq:A1Dn-Higgs} and not turn on any of the $L_i$ and $\tilde L^i$ vevs.} The class $\CS$ interpretation of this flow is that we are Higgsing the regular singularity of the $(A_1, D_{2n})$ theory. As a result, we expect to have a decoupled axion-dilaton multiplet combined with the $(A_1, A_{2n-3})$ theory coming from the irregular singularity. To see this flow somewhat more explicitly, we again consider the $S^1$ reduction. The resulting long-distance SCFT can alternatively be described as the IR fixed point of the three-dimensional $\CN=4$ abelian gauge theory whose matter content is summarized in Table \ref{AAsideD2n} (note that when we reduce the $(A_1, D_{2n})$ theory on $S^1$, we never flow to the free limit of the gauge theory).\foot{Our above description can be derived by applying mirror symmetry to the three-dimensional construction given in \rcite{Xie:2012hs, Boalch:XXXXxx, Boalch:YYYYyy}.}
\begin{table}
\begin{center}
     \begin{tabular}{| c | c | c | c | c | c |}
\hline   & $U(1)_1$ & $U(1)_2$ & $\cdots$ & $U(1)_{n-2}$& $U(1)_{n-1}$\cr\hline\hline
        $q_1$ & $1$ & $0$& $\cdots$ & 0 & 0\cr\hline
        $q_2$ & $1$ & $0$& $\cdots$ & 0 & 0\cr\hline
        $q_3$ & $1$ & $1$& $\cdots$ & 0 & 0\cr\hline
        $\vdots$ & $\vdots$ & $\vdots$& $\ddots$ & $\vdots$ & \vdots\cr\hline
        $q_n$ & $0$ & $0$& $\cdots$ & 1 & 1\cr\hline
        $q_{n+1}$ & $0$ & $0$& $\cdots$ & 0 & 1\cr\hline
      \end{tabular}
\caption{The matter fields (and charges) of the three-dimensional gauge theory that flows to the $S^1$ reduction of the $(A_1,D_{2n})$ theory.}
\label{AAsideD2n}
\end{center}
\end{table}
This theory has the following superpotential
\eqn{
W=\Phi_1(q_1\tilde q^1+q_2\tilde q^2+q_3\tilde q^3)+\sum_{a=2}^{n-1}\Phi_a(q_{a+1}\tilde q^{a+1}+q_{a+2}\tilde q^{a+2})~.
}[A1D2msuperpot]
We can think of $(q_{1,2}, \tilde q^{\dagger}_{1,2})$ as comprising the \lq\lq fields of the regular singularity," (they transform as doublets under the $SU(2)$ flavor symmetry), while the remaining fields can be thought of as the \lq\lq fields of the irregular singularity" (we can think of the $U(1)$ symmetry as giving $q_3$ charge $-1$, $\tilde q^3$ charge $+1$, and leaving the rest of the fields neutral).\foot{In the case of $n=2$, we see that the $q_i$ form a triplet of $SU(3)$.} We expect the theory described in Table \ref{AAsideD2n} to capture the Higgs branch of the original AD SCFT. Indeed, we have the following operator maps
\eqn{
M_i^{\ j}\leftrightarrow q_i\tilde q^j~, \ \ \ L_i\leftrightarrow q_i \tilde q^3 q_4\cdots \tilde q^{n+1}~, \ \ \ \tilde L^i\leftrightarrow \tilde q^iq_3\tilde q^4\cdots q_{n+1}~,
}[HiggsBrMapA1D2n]
where $i,j =1,2$.\foot{In writing \eqref{HiggsBrMapA1D2n}, we are assuming $n$ is even. In the case of odd $n$, we should replace $\tilde q^{n+1}\leftrightarrow q_{n+1}$.} 

Turning on $\langle M_1^{ \ 2}\rangle\ne0$ has the effect of breaking the flavor symmetry from $SU(2)\times U(1)\to U(1)$ (although, in the deep IR, we find an emergent $Sp(1)$ flavor symmetry associated with the axion-dilaton). Moreover, we see from Table \ref{AAsideD2n} that the $U(1)_1$ gauge symmetry is Higgsed, leaving one linear combination of the fields of the regular singularity to play the role of the axion-dilaton. The remaining theory is a $U(1)^{n-2}$ gauge theory with $n-1$ matter fields $(q_{k}, \tilde q^{\dagger}_k)$ ($k=3,\cdots, n+1$) and superpotential
\eqn{
W=\sum_{a=2}^{n-1}\Phi_a(q_{a+1}\tilde q^{a+1}+q_{a+2}\tilde q^{a+2})~.
}[redW]
In other words, in addition to the decoupled axion-dilaton, we find the $S^1$ reduction of the $(A_1, A_{2n-3})$ theory at long distance.

From the perspective of the index, compatibility with \eqref{GRR} then requires
\eqn{
\CI_{\rm vect}^{-1}\cdot\CI_{(A_1, A_{2n-3})}(q;x)=-2{\rm Res}_{y=q^{-{1\over2}}}\left({1\over y}\CI_{(A_1, D_{2n})}(q;x,y)\right)~.
}[GRRAtoD]
To verify this equation, it is useful to rewrite our conjecture in \eqref{eq:A1Dn-index} as follows 
\begin{eqnarray}\label{A1D2nindex}
\CI_{(A_1, D_{2n})}(q; x, y)&=&{\left(\CI_{\rm vect}^{SU(2)}(y)\right)^{-{1\over2}}\over(q;q)_{\infty}}\sum_{k=0}^{\infty}\Big({1\over y-y^{-1}}\sum_{s=\pm1}\left[{q^{n(2k+1)^2-2nk^2\over2}y^{2k+2}x^s\over1-q^{n(k+{1\over2})}yx^s}-{q^{n(2k+1)^2-2nk^2\over2}y^{-2k-2}x^s\over1-q^{n(k+{1\over2})}y^{-1}x^s}\right]\nonumber\\ &+&q^{nk(k+1)}\chi_{2k+1}(y)\Big)~.
\end{eqnarray}
We see that the pole at $y=q^{-{1\over2}}$ is manifest when we rewrite $\left(\CI_{\rm vect}^{SU(2)}(y)\right)^{-{1\over2}}$ as
\eqn{
\left(\CI_{\rm vect}^{SU(2)}(y)\right)^{-{1\over2}}={1\over1-q y^{2}}\cdot{1\over(q;q)_{\infty}}{1\over1-qy^{-2}}\prod_{m=2}^{\infty}{1\over(1-q^my^2)(1-q^my^{-2})}~.
}[VectIndexrewrite]
Plugging \eqref{VectIndexrewrite} into \eqref{A1D2nindex} and taking the residue at $y=q^{1\over2}$ we recover \eqref{GRRAtoD}. 
This result is a non-trivial check of our conjectures \eqref{eq:A1An-index} and \eqref{eq:A1Dn-index}.\footnote{To be precise, this result follows from the structure of \eqref{eq:A1An-index}, \eqref{eq:A1Dn-index}, and the expression for $f_{R}(q;x)$ given in \eqref{eq:trace-regular}. In particular, it is independent of the expression for $\tilde{f}_R^{(n)}(q;x)$ given in \eqref{eq:proposal}.}

\subsubsection{The $(A_1, D_{2n})\rightarrow (A_1, A_1)\oplus(A_1, A_1)$ RG flow}

Finally, we can consider the RG flow
\eqn{
(A_1, D_{2n})\rightarrow (A_1, A_1)\oplus(A_1, A_1)~.
}[DtoEmpty]
We claim this RG flow can be initiated by, for example, turning on $\langle\tilde L^2\rangle\ne0$ and keeping the rest of the vevs for the operators in Table \ref{table:A1Dn-Higgs} vanishing (this motion on the moduli space breaks $SU(2)\times U(1)\to U(1)$ and $SU(2)_R\to SO(2)_R$).\foot{We can also initiate this flow by turning on vevs of other operators so long as we turn on the vev for at least one of the $L_i$ or $\tilde L^j$ and satisfy \eqref{eq:A1Dn-Higgs}.} From the class $\CS$ perspective, one potential interpretation of the flow in \eqref{GRRfinal} is that it corresponds to Higgsing both the irregular and regular singularities down to a diagonal singularity (which is itself irregular). It would be interesting to make these notions more precise.

From the $S^1$ reduction described in Table \ref{AAsideD2n} and the operator maps in \eqref{HiggsBrMapA1D2n}, we see that turning on this vev completely Higgses the three-dimensional $U(1)^{n-1}$ gauge theory. In the prescription of \eqref{GRR}, we should study the pole at $x=yq^{n\over2}$. More precisely, compatibility with \eqref{DtoEmpty} requires
\eqn{
\CI_{\rm vect}^{-1}\cdot \CI_{(A_1, A_1)}={\rm Res}_{x=yq^{n\over2}}\left({1\over x}\CI_{(A_1, D_{2n})}(q;x,y)\right)~.
}[GRRfinal]
Identifying $\mathfrak{y}\equiv q^{1\over2}y^2$, we see that indeed this relation holds with $\CI_{(A_1, A_1)}=\prod_{m=0}^{\infty}(1-q^{m+{1\over2}}\mathfrak{y})^{-1}(1-q^{m+{1\over2}}\mathfrak{y}^{-1})^{-1}$, which is the usual Schur index for the free hypermultiplet.

Let us explain why it is natural to shift the fugacity by taking $\mathfrak{y}\equiv q^{1\over2}y^2$. From the perspective of the four-dimensional theory this shift is somewhat mysterious since setting $x=yq^{n\over2}$ in \eqref{GRRfinal} only implies that we should take $R\to R+{n\over2}f$ and $J_3\to J_3+{f\over2}$ (where $f$ is the $U(1)$ charge) in the IR. To understand the shift $y^2\to\mathfrak{y}q^{-{1\over2}}$ more naturally, we note that it implies $R\to R+{n\over2}f-{1\over2}(J_3+{1\over2}f)$. Under the assumption that the IR $SU(2)_R$ Cartan is visible in the UV, this $R$ symmetry mixing with $J_3+{1\over2}f$ is the only such mixing that allows a pair of chiral Higgs branch operators of the UV SCFT to flow to operators in the IR with $SU(2)_R$ weight $1/2$ (in this case, the $\tilde L^1$ and $M_1^{\ 2}$ operators). These operators are then natural candidates for the UV ancestors of the free IR SCFT fields.\foot{For another explanation of the shift $\mathfrak{y}\equiv q^{1\over2}y^2$,  it is useful to recall the three-dimensional theory described in Table \ref{AAsideD2n}. There, turning on the vev $\langle \tilde L^2\rangle\ne0$ leads to a flow to an IR theory involving a decoupled axion dilaton and an IR SCFT comprising of the free $q_1$ and $\tilde q^1$ fields. These latter fields should have $SU(2)_R$ weight $1/2$. Moreover, the three-dimensional fields in the product $\tilde q^2q_3\cdots$ that maps to the $\tilde L^2$ operator in \eqref{HiggsBrMapA1D2n} should have vanishing $R$-charge. These constraints are sufficient to fix the three-dimensional $SU(2)_R$ Cartan that is preserved along the flow as $R\to R+{2n-1\over4}f-{1\over2}J_3+{1\over4}U(1)_1+\sum_{m=2}^{n-1}(-1)^m\left(n-m\over2\right)U(1)_m$ (where the $U(1)_i$ in this expression are the global gauge symmetries in three dimensions). The coefficient of $f$ in this shift precisely fixes the exponent of $q$ in $y^2\to\mathfrak{y}q^{-{1\over2}}$ (the exponent of $\mathfrak{y}$ is simply a matter of convention).} This result is another important check of our conjecture.

\subsection{Cardy-like behavior}
There is a large body of evidence (e.g., \rcite{Imamura:2011uw, Spiridonov:2012ww, Aharony:2013dha, Ardehali:2014zba, DiPietro:2014bca, Ardehali:2015hya}) suggesting that the small $S^1$ limit of the index behaves, in some ways, analogously to the high-temperature limit of the partition function of two-dimensional CFTs described by Cardy \rcite{Cardy:1986ie}. This Cardy-like behavior of the index is governed by the linear combination of central charges $a-c$. More precisely, the authors of \rcite{DiPietro:2014bca} argued that in a large class of theories, the leading behavior of the index (or $S^1\times S^3$ partition function) in the limit of small $S^1$ is given by
\eqn{
\lim_{\beta\to0}\log\CI=-{16\pi^2\over3\beta}(a-c)+\cdots=-{\pi^2\over3\beta}{\rm Tr} R_{\CN=1}+\cdots~,
}[CardyBehavior]
where $\beta$ is the circumference of the $S^1$ (we set the radius of the $S^3$ to unity and consider the round three-sphere), $R_{\CN=1}$ is the $\CN=1$ superconformal $R$ symmetry (it assigns charge $2/3$ to free chiral multiplets), and the ellipses contain sub-leading terms in the small $\beta$ limit.\foot{One may also have contributions from linear abelian flavor anomalies, ${\rm Tr} G$, at $\CO(\beta^{-1})$ in \eqref{CardyBehavior} \rcite{DiPietro:2014bca} which have the effect of shifting the $R$-symmetry. However, on general grounds, such anomalies vanish for $\CN=2$ flavor symmetries (but {\it not} for the linear combination of $U(1)_R$ and Cartan of $SU(2)_R$ that is a flavor symmetry from the $\CN=1\subset\CN=2$ point of view) \rcite{Buican:2013ica}.}

In this section, we will check the consistency of our conjecture for the index with this type of Cardy-like behavior. However, we should give a word of caution. The $\CN=2$ superconformal index has many different limits. Some of these limits are subtle because contributing operators are annihilated by many different $\CN=2$ supercharges. In particular, this fact sometimes implies the absence of derivative contributions (depending on the preserved supercharges) and hence the absence of Cardy-like behavior (by which we mean that the index does not have an essential singularity in $\beta$; note that this does not contradict the statement in \eqref{CardyBehavior}). Examples of such limits include both the Coulomb branch and Hall-Littlewood limits described in \rcite{Gadde:2011uv}.\foot{However, it is interesting to note that the Coulomb branch limit captures a different linear combination of the conformal anomalies, $2a-c$ \rcite{Buican:2014qla}.} On the other hand, the Schur limit receives contributions from certain space-time derivatives, and so it is reasonable to believe that it should exhibit Cardy-like behavior.

\subsubsec{Cardy-like behavior of previously known indices}
Before checking the compatibility of our conjecture with this discussion, let us first illustrate the Cardy-like behavior of the $\beta\to0$ limit of the Schur index in a series of increasingly complicated theories whose indices are already known. To that end, let us first consider the free hypermultiplet. The log of its Schur index is \rcite{Gadde:2011uv}
\begin{equation}\label{hyperlogI}
\log\CI_{H}=\sum_{n=1}^{\infty}{q^{n\over2}\over n(1-q^n)}(x^n+x^{-n})~,
\end{equation}
where  $q=e^{-\beta}$ (and $x$ is a flavor fugacity). Taking the limit $\beta\to0$, we find
\begin{equation}\label{hhcont}
\lim_{\beta\to0}\log\CI_{H}={2\over\beta}\sum_{n=1}^{\infty}n^{-2}+\cdots={\pi^2\over3\beta}+\cdots=-{\pi^2\over2\beta}{\rm Tr} R_{\CN=1}+\cdots~.
\end{equation}
Note that while this limit of the $\CN=2$ index has Cardy-like behavior, the precise coefficient of $\beta^{-1}$ differs by a factor of $3/2$ relative to \eqref{CardyBehavior}. This difference is not surprising since, in our limit of the index, states contribute only if they are annihilated by $\tilde\CQ_{2\dot-}$ and $\CQ^1_{+}$ (recall that we are using the supercharge conventions of \rcite{Buican:2014qla}). Next, let us consider the free $\CN=2$ vector multiplet. We have 
\begin{equation}
\log\CI_{V}=-2\sum_{n=1}^{\infty}{q^n\over n(1-q^n)}~.
\end{equation}
In the small $S^1$ limit, we have
\begin{equation}\label{vectorlogI}
\lim_{\beta\to0}\log\CI_{V}=-{2\over\beta}\sum_{n=1}^{\infty}n^{-2}+\cdots=-{\pi^2\over3\beta}+\cdots=-{\pi^2\over2\beta}{\rm Tr} R_{\CN=1}+\cdots~,
\end{equation}
which is again consistent with Cardy-like behavior.

Let us now consider interacting $\CN=2$ theories. One important set of such theories are Gaiotto's $T_N$ theories (with $N\ge3$), which we alluded to in the introduction \rcite{Gaiotto:2009we}. In the class $\CS$ context described in section \ref{sec:AD_from_6d}, they are engineered by taking the $\mathfrak{g}=A_{N-1}$ $(2,0)$ theory on a sphere, $\CC={\bf P}^1$, with three regular punctures.\foot{For $N>2$, these punctures carry additional data. In particular, for the $T_N$ case, they are so-called \lq\lq full" punctures.} The Schur index of these theories was originally conjectured in \rcite{Gadde:2011ik} 
\eqn{
\CI_{T_N}={(q;q)^{N-1}_{\infty}{\prod_{i=1}^3}P.E.\left[{q\over1-q}\chi_{\rm adj}^{su(N)}(x_i)\right]\over\prod_{\ell=1}^{N-1}(1-q^{\ell})^{N-\ell}}\sum_{R}{1\over[{\rm dim}\ R]_q}\chi_{\CR}^{su(N)}(x_1)\chi_{\CR}^{su(N)}(x_2)\chi_{\CR}^{su(N)}(x_3)~,
}[TNindex]
where $R$ now runs over the irreducible representations of $su(N)$, the $x_i$ are flavor fugacities, and
\eqn{
[{\rm dim}\ R]_q=\prod_{i<j}{[\lambda_i-\lambda_j+j-i]_q\over[j-i]_q}~.
}[defRq]
In \eqref{defRq}, $\lambda_1\ge\lambda_2\ge\cdots\ge\lambda_{N-1}\ge\lambda_N=0$ are integers describing the  lengths of the rows of boxes comprising the Young tableau corresponding to $R$. It is then straightforward to compute the leading behavior in the small $\beta$ limit of \eqref{TNindex}. Indeed, the contributions at leading order in $\beta$ come exclusively from the $(q;q)_{\infty}^{N-1}$ and ${\prod_{i=1}^3}P.E.\left[{q\over1-q}\chi_{\rm adj}^{su(N)}(x_i)\right]$ factors. The first of these factors gives $(N-1)/2$ times the vector multiplet contribution in \eqref{vectorlogI}, while the second of these factors gives $-3(N^2-1)/2$ times the vector multiplet contribution. As a result, we have
\eqn{
\lim_{\beta\to0}\log\CI_{T_N}={(3N+2)(N-1)\over2}\cdot{\pi^2\over3\beta}+\cdots=-{\pi^2\over2\beta}{\rm Tr}R_{\CN=1}+\cdots~,
}[TnIndex]
which agrees with the computation of the central charges in \rcite{Gaiotto:2009gz} (it is also easy to see that, in the case of $N=2$, i.e., the free $T_2$ theory, our result equals the answer for eight half-hypermultiplets).

\subsubsec{Cardy-like behavior of the $(A_1, A_{2n-3})$ and $(A_1, D_{2n})$ indices}
Let us now apply the above discussion and examine the leading divergences as $\beta\to0$ for our conjectured AD indices. From \eqref{acA1A2m3} and \eqref{acA1D2n}, we see that $a-c$ for the $(A_1, A_{2n-3})$ theory is the same as for one hypermultiplet, while $a-c$ for the $(A_1, D_{2n})$ theory is the same as for two hypermultiplets. It is straightforward to check that our conjectured forms of the indices then exhibit the expected Cardy-like behavior. 

Indeed, in the case of the $(A_1, A_{2n-3})$ theory, we have a leading contribution from the $\CN(q)$ factor that is equal to a half-hypermultiplet contribution and a similar contribution from the irregular singularity wave function \eqref{eq:proposal}. As a result, we find
\eqn{
\lim_{\beta\to0}\CI_{(A_1, A_{2n-3})}={2\pi^2\over3\beta}+\cdots=-{\pi^2\over2\beta}{\rm Tr}R_{\CN=1}+\cdots~.
}[A1A2n3Cardy]
For the $(A_1, D_{2n})$ theory, we find a leading contribution equal to that of a half-hypermultiplet from the irregular singularity wave function \eqref{eq:proposal} and a leading contribution from the regular singularity wave function \eqref{eq:trace-regular} equal to that of three half-hypermultiplets. As a result, we find
\eqn{
\lim_{\beta\to0}\CI_{(A_1, D_{2n})}={4\pi^2\over3\beta}+\cdots=-{\pi^2\over2\beta}{\rm Tr}R_{\CN=1}+\cdots~.
}[A1D2nCardy]
The results in \eqref{A1A2n3Cardy} and \eqref{A1D2nCardy} are non-perturbative in $q$ checks of our conjectures (note that they are compatible with the results for the free hyper, free vector, and $T_N$ theories in \eqref{hhcont}, \eqref{vectorlogI}, and \eqref{TnIndex}).\foot{Note that we are using the notation \lq\lq ${\rm Tr} R_{\CN=1}$" in \eqref{A1A2n3Cardy} and \eqref{A1D2nCardy} rather loosely since the AD theory is non-Lagrangian (and we cannot write the linear $R_{\CN=1}$ anomaly as a sum over charges of weakly coupled fields; in the case of the non-Lagrangian $T_N$ theories we can give a more transparent meaning to this trace by use of generalized Argyres-Seiberg duality). Instead, we use this term to stand for the appropriate contact term in the correlation function of the $R$ current and two stress tensors.}

\section{Discussion and Conclusions}

In this paper, we have proposed closed-form expressions for the Schur indices of the $(A_1,A_{2n-3})$ and $(A_1,D_{2n})$ theories. Our proposal is motivated by the relation between the Schur index and two-dimensional $q$-deformed YM theory found in \rcite{Gadde:2011ik}. We have performed various consistency checks of our conjectures ranging from compatibility with certain two-dimensional chiral algebras via the framework in \rcite{Beem:2013sza} (in the cases of the $(A_1, A_3)$ and $(A_1, D_4)$ SCFTs) to compatibility with an intricate set of RG flows. We have also checked that out conjectures reproduce the $S$-duality recently discussed in \rcite{Buican:2014hfa}, that they respect known Higgs branch relations, and that they are compatible with expected Cardy-like behavior in the limit of small $S^1$ radius.

Moreover, our work suggests interesting ways in which AD theories behave as particularly simple $\CN=2$ SCFTs. In the introduction we discussed a manifestation of this idea by studying the scaling behavior of the $a$ anomaly with respect to the dimension of the Coulomb branch in \eqref{aVsRank}. Furthermore, in section 4.2 we saw hints of another, complementary, way in which AD theories may be simple: given some AD theory with a non-abelian flavor symmetry, $\CT$, the corresponding chiral algebra, $\chi_{\CT}$, should not have too many AKM modules or else we may find logarithmic correlation functions (in violation of physical expectations in four dimensions).

There are many interesting directions for future work. Here we only list a few:

\begin{itemize}
\item Make the notion of simplicity discussed in the previous paragraph more rigorous.\foot{In an upcoming paper we will see yet another very different manifestation of this idea \rcite{Buican:2015hsa}.}

\item  Generalize our conjectures for the Schur limit to the full superconformal index. Doing so would provide us with more refined information about the operator spectrum of the $(A_1,A_{2n-3})$ and $(A_1,D_{2n})$ theories.

\item Generalize our results to other AD theories. We have only studied AD theories descending from the six-dimensional $A_1$ theory. For example, there are AD theories obtained by compactifying the 6d $(2,0)\, A_{m}$ theory. Studying these more general theories would also allow us to further explore the landscape of AD theories with exactly marginal deformations and perhaps better understand the resulting conformal dualities \rcite{Buican:2014hfa} (see also the largely complementary recent work \rcite{DelZotto:2015rca}). Note that we already discussed one of the simplest such generalizations in \eqref{eq:A3A3-integral}. However, we treated this case perturbatively in the various building blocks coming from the 6d $(2,0)$ $A_1$ theory. Therefore, even in this case, it would be nice to have a non-perturbative in $q$ expression.

\item Identify the two-dimensional chiral algebras, $\chi_{(A_1, A_{2n-3})}$ and $\chi_{(A_1, D_{2n})}$, associated with the $(A_1,A_{2n-3})$ and $(A_1,D_{2n})$ theories (in the sense of \rcite{Beem:2013sza}) for general $n$. In particular, for the $(A_1,D_{2n})$ theory, the chiral algebra contains the Virasoro algebra with central charge $c_{2d} = -6n+4$ and the affine $su(2)\times u(1)$ algebra at level $k_{2d}=-\frac{2n-1}{2}$. Note that $c_{2d}$ is precisely the Sugawara central charge for the affine $su(2)\times u(1)$ algebra. Therefore, we expect that, for $n>2$, the two-dimensional stress tensor is given by the Sugawara stress tensor of the affine algebra.\footnote{In the case $n=2$, as discussed in sub-section \ref{subsec:2d-chiral}, the stress tensor is given by the Sugawara stress tensor for $\widehat{su}(3)_{-\frac{3}{2}}$.} However, our formula for the Schur index implies that the two-dimensional chiral algebra contains more than the affine $su(2)\times u(1)$ algebra.\foot{This is analogous to the situation of the $\CN=2,\,SU(N)$ gauge theory with $N_f=2N$ flavors for $N>2$ studied in \rcite{Beem:2013sza}.} It would be very interesting to identify the full set of generators of this chiral algebra (we should at least include the baryons described in section 5).\foot{There is a conjecture for the $(A_1, A_{2n-3})$ chiral algebra \rcite{conjecture}.}

\item Further study the state, $|x;n\rangle$, corresponding to an irregular puncture. In the AGT relation \rcite{Alday:2009aq} between the $S^4$ partition function and Liouville theory, the state corresponding to an irregular puncture can be naturally constructed in a ``colliding limit'' of states corresponding to regular punctures \rcite{Gaiotto:2012sf} (see also \rcite{Nishinaka:2012kn, Choi:2014qha, Rim:2015tsa} for its matrix model description). It is therefore interesting to look for a similarly natural construction of $|x;n\rangle$ from the state corresponding to a regular puncture in the 2d $q$-deformed YM theory.

\item Finally, it would be interesting to see if the large $n$ limit of our results for the index sheds light on possible gravitational duals~for~the~$(A_1, A_{2n-3})$~and~$(A_1, D_{2n})$~theories~(note that $a/c\to1$ as $n\to\infty$ in \eqref{acA1A2m3} and \eqref{acA1D2n}).

\end{itemize}

\ack{ \bigskip
We are grateful to L.~Di~Pietro, Z.~Komargodski, D.~Kutasov, C.~Papageorgakis, L.~Rastelli, N.~Seiberg, and Y.~Tachikawa for interesting discussions and communications. We would like to particularly thank Y.~Tachikawa for drawing our attention to the implications of \rcite{Beem:2013sza} for the Schur indices of the $(A_1, A_3)$ and  $(A_1, D_4)$ theories when one of us (T.~N.) gave a talk at the University of Tokyo. We have also benefitted from discussions and collaborations with  G.~Moore, C.~Papageorgakis, and D.~Shih on closely related topics. M.~B. would like to thank the members of the Queen Mary University of London CRST and the members of the University of Chicago EFI for stimulating scientific environments and discussions while parts of this work were being completed. Our research is partially supported by the U.S. Department of Energy under grants DOE-SC0010008, DOE-ARRA-SC0003883, and DOE-DE-SC0007897.
}

\newpage
\begin{appendices}
\section{$q$-deformed two-dimensional YM theory}
\label{app:qYM}

In this appendix, we review an expression for the partition function of $q$-deformed two-dimensional $SU(2)$ YM theory on a Riemann surface, $\CC_{g,m}$, of genus $g$ with $m$ punctures (for more details, the interested reader can consult \rcite{Cordes:1994fc}). To proceed, first note that when we calculate the partition function, we have to specify the boundary condition at each puncture. This data corresponds to an ``external state'' on a small $S^1$ surrounding the puncture.

In the case of 2d $SU(2)$ YM theory, the Hilbert space of states, $\mathcal{H}$, is known to be the space of $SU(2)$ class functions \rcite{Witten:1991we, Minahan:1993np}. This space has a natural orthonormal basis, $|R\rangle$, labeled by the irreducible representations, $R$, of $SU(2)$. On the other hand, $\mathcal{H}$ has another basis, $|U\rangle = \sum_R \chi_R^{su(2)}(U)|R\rangle$, for $U\in SU(2)$, which is called the ``holonomy basis.'' Indeed, $|U\rangle$ corresponds to the boundary condition with fixed holonomy $U$ around a puncture. Since $|U\rangle = |VUV^{-1}\rangle$ for all $V\in SU(2)$, the state $|U\rangle$ depends only on the conjugacy class of $U$. In what follows, we denote $|U\rangle$ by $|x\rangle_0$ with $x,x^{-1}\in U(1)$ being the eigenvalues of $U$.

The partition function of the theory on $\mathcal{C}_{g,m}$ is called the $m$-point function, and it maps an external state $|\psi\rangle \in \mathcal{H}^{\otimes m}$ to ${\bf C}$. Therefore, the Riemann surface, $\mathcal{C}_{g,m}$, is associated with a vector, $\langle \mathcal{C}_{g,m}|\in (\mathcal{H}^*)^{\otimes m}$, and the $m$-point function is given by $Z_{\text{YM}}(\psi) = \langle \mathcal{C}_{g,m}|\psi\rangle$. Moreover, since two-dimensional YM theory is invariant under area-preserving diffeomorphisms, it becomes a TQFT (in the sense of \rcite{Atiyah:1989vu}) in the limit we take the area of $\mathcal{C}_{g,m}$ to zero.\footnote{To be precise, the Hilbert space of states in \rcite{Atiyah:1989vu} is finite dimensional while ours is infinite dimensional. This difference, however, is not important for our purposes.}  This fact implies that $\langle\mathcal{C}_{g,m}|$ is fixed by the vectors $\langle \mathcal{C}_{0,1}|,\,\langle \mathcal{C}_{0,2}|,\, \langle \mathcal{C}_{0,3}|$ (these are the vectors for the disk, the cylinder, and the pair of pants respectively).

However, we are interested in the $q$-deformed theory. This version is obtained by deforming $\langle \mathcal{C}_{g,m}|$ with one parameter $q\in {\bf C}$ \rcite{Aganagic:2004js, Buffenoir:1994fh,Klimcik:1999kg} (we recover ordinary two-dimensional YM theory by setting $q=1$). In the zero-area limit, the deformed theory generally also becomes a TQFT. The vectors $\langle \mathcal{C}_{g,m}|$ for the disk, the cylinder, and the pair of pants are now given by
\begin{align}
\langle \text{disk}| &= \sum_{R} [\text{dim}\,R]_q\langle R|~,\ \ \  \langle \text{cylinder} | = \sum_{R}\langle R| \otimes \langle R|~,
\nonumber\\
\langle \text{pants} | &= \sum_{R}\frac{1}{[\text{dim}\, R]_q}\langle R| \otimes \langle R| \otimes \langle R|~,
\end{align}
where $[k]_q = (q^{\frac{k}{2}}-q^{-\frac{k}{2}})/(q^{\frac{1}{2}}-q^{-\frac{1}{2}})$. 
The vectors for all the other Riemann surfaces are obtained by ``gluing'' these three basic vectors, where the gluing procedure is given by the natural map $\langle R| \otimes \langle R'| \to \langle R|R'\rangle = \delta_{R,R'}$.
Therefore, a general Riemann surface, $\mathcal{C}_{g,m}$, is associated with the state
\begin{align}
\langle \mathcal{C}_{g,m}| = \sum_{R} \big([\text{dim}\,R]_q\big)^{2-2g-m}\langle R| \otimes \langle R| \otimes \cdots \otimes \langle R|~.
\label{eq:C-vector}
\end{align}

Now, suppose that we fix the holonomy of the gauge field around the punctures on $\mathcal{C}_{g,m}$, with the external state given by $|\vec{x}\rangle_0 \equiv |x_1\rangle_0\otimes \cdots \otimes |x_m\rangle_0 \in \mathcal{H}^{\otimes m}$ for $x_i\in U(1)$.
The partition function of $q$-deformed YM theory on $\mathcal{C}_{g,m}$ in the zero-area limit is then given by 
\begin{align}
Z_{\text{qYM}}(x_1,\cdots,x_m) = \sum_{R}\left([\text{dim}\,R]_q\right)^{2-2g-m}\prod_{k=1}^{m}\chi_R^{su(2)}(x_k)~.
\end{align}
Here the sum over $R$ stands for the sum over intermediate states, while $\chi_R^{su(2)}(x_k)$ is the inner product, $\langle R|x_k\rangle_0$, of an intermediate state, $|R\rangle$, and the external state $|x_k\rangle_0$.

In the main text, we study the superconformal index of class $\mathcal{S}$ theories. If all the punctures on $\mathcal{C}_{g,m}$ are associated with regular defects, the Schur index of $\mathcal{T}_{\mathcal{C}_{g,m}}$ is given by \eqref{eq:regular}, which is identical to $Z_{\text{qYM}}(x_1,\cdots,x_m)$ up to a prefactor. Moreover, the prefactor can be absorbed by rescaling the basic vectors of the $q$-deformed YM theory as
$\langle \text{disk}| \to \mathcal{N}(q)\langle \text{disk}|,\; \langle\text{pants}| \to  \frac{1}{\mathcal{N}(q)}\langle \text{pants}|$, and $|x\rangle_0 \to \eta^{\frac{1}{2}}(x)|x\rangle_0$ (where $\eta^{1\over2}(x)\equiv P.E.\left({q\over1-q}\chi_{\rm adj}^{su(2)}(x)\right)$). 
In particular, we see that all the ingredients depending on $x_k$ are now included in
\begin{align}
|x_k\rangle \equiv \eta^{\frac{1}{2}}(x_k)|x_k\rangle_0~.
\end{align}
Thus, the $k$-th regular puncture on $\mathcal{C}_{g,m}$ maps to the vector
$|x_k\rangle\in \mathcal{H}$, which is uniquely determined by the inner product \eqref{eq:inner-regular}.

\section{The characters of affine Lie algebras at negative levels}
\label{app:character}

In this appendix we briefly review a formula \rcite{Kac-Wakimoto} for evaluating the characters of irreducible highest weight modules of affine $A_1$ and $A_2$ algebras with negative levels. To that end, let $\mathfrak{g}$ be the affine Lie algebra associated with a finite dimensional simple Lie algebra, $\overline{\mathfrak{g}}$. We denote the level of $\mathfrak{g}$ by $k$. We are particularly interested in the cases $(\bar{\mathfrak{g}},k) = (A_1,-\frac{4}{3})$ and $(\bar{\mathfrak{g}},k) = (A_2,-\frac{3}{2})$. Let $\Delta_\pm$ be the sets of positive and negative roots of $\mathfrak{g}$ and $\bar{\Delta}_\pm$ be those of $\overline{\mathfrak{g}}$. We also define $\bar{\Delta} = \bar{\Delta}_+ \cup \bar{\Delta}_-$. Then $\Delta_\pm = \{\bar{\alpha}+n\delta|\bar{\alpha}\in \bar{\Delta},n\in{\bf Z}_\pm \}\cup \{\bar{\alpha}\in \bar{\Delta}_\pm\}\cup \{n\delta|n\in{\bf Z}_{\pm}\}$, where $\delta =  \sum_{i=0}^{\text{rank}\,\overline{\mathfrak{g}}} a_i \alpha_i$ with $a_i$ and $\alpha_i$ being the marks and the simple roots of $\mathfrak{g}$, respectively.
We denote by $\mathfrak{g}_\alpha$ the root space for a root $\alpha\in \Delta_+\cup \Delta_-$, and define $\mathfrak{n}_\pm = \bigoplus_{\alpha\in\Delta^\pm}\mathfrak{g}_\alpha$. Then $\mathfrak{g}$ has the following root decomposition:
\begin{align}
\mathfrak{g} = \mathfrak{h} \oplus \mathfrak{n}_+ \oplus \mathfrak{n}_-~,
\end{align}
where $\mathfrak{h}$ is the Cartan subalgebra of $\mathfrak{g}$.
It follows that $\text{dim}\,\mathfrak{g}_{n\delta} = \text{rank}\, \overline{\mathfrak{g}}$ for $n\in{\bf Z}\setminus\{0\}$ and  $\text{dim}\,\mathfrak{g}_{\bar{\alpha} + n\delta} = 1$ for $\bar{\alpha}\in\bar{\Delta}$ and $n\in{\bf Z}$.

A highest weight module of $\mathfrak{g}$ with highest weight $\lambda\in \mathfrak{h}^*$ is a $\mathfrak{g}$-module, $V$, with $v_\lambda \in V$ such that (i) $hv_\lambda = \lambda(h)v_\lambda$ for all $h\in\mathfrak{h}$, (ii) $x v_\lambda=0$ for all $x\in\mathfrak{n}_+$, and (iii) $V=U(\mathfrak{g})v_\lambda$. Here $U(\mathfrak{g})$ is the universal enveloping algebra of $\mathfrak{g}$. A highest weight module $V$ has a weight decomposition $V=\oplus_\mu V_\mu$ such that $hv=\mu(h)v$ for all $h\in \mathfrak{h}$ and $v\in V_\mu$.
Its character is defined by
\begin{align}
\text{ch}\, V = \sum_{\mu\in \mathfrak{h}^*}(\text{dim} V_\mu)e^{\mu}~,
\end{align}
where $e^\mu$ is a formal exponential such that $e^{\mu_1} e^{\mu_2} = e^{\mu_1+\mu_2}$.
The largest highest weight module is called the Verma module, which is isomorphic to $U(\mathfrak{n}_-)$ as a vector space. Let $\{x_i|i=1,2,3,\cdots\}$ be an ordered basis of $\mathfrak{n}_-$. Then, the Poincar\'e-Birkhoff-Witt theorem implies that the Verma module $M(\lambda)$ is spanned by $(x_1^{n_1}x_2^{n_2}x_3^{n_3}\cdots )v_\lambda$ with $n_i\in{\bf N}$ vanishing except for finite number of $i$. Therefore, the character of the Verma module is of the form
\begin{align}
\text{ch}\, M(\lambda) = e^{\lambda}\prod_{n=1}^\infty \left[\frac{1}{(1-q^n)^{\text{rank}\,\bar{\mathfrak{g}}}}\prod_{\bar{\alpha} \in \bar{\Delta}_+}\frac{1}{(1-q^n e^{\bar{\alpha}})(1-q^{n-1}e^{-\bar{\alpha}})}\right]~,
\end{align}
where we define $q = e^{-\delta}$.

Note that the Verma module is not always irreducible for general $\lambda$. The irreducible highest weight module, $L(\lambda)$, is defined as the quotient of $M(\lambda)$ by its maximal proper submodule. The character of $L(\lambda)$ is generally written as
\begin{align}
\text{ch}\,L(\lambda) = \sum_{\mu \in \mathfrak{h}^*} m_{\lambda,\mu}\; \text{ch}\,M(\mu)~,
\end{align}
with $m_{\lambda,\mu}\in {\bf Z}$. If the highest weight $\lambda$ is integral and dominant, namely if the Dynkin labels of $\lambda$ are all non-negative integers, then  $m_{\lambda,\mu}$ are easily evaluated via the famous Weyl-Kac formula.

In our cases, however, $\lambda$ is neither integral nor dominant. 
Therefore we need a generalization of the Weyl-Kac formula, which we refer to as the ``Kac-Wakimoto formula.''\footnote{In \rcite{Beem:2013sza}, the authors used the Kazhdan-Lusztig formula to evaluate the characters of $\widehat{so}(8)_{}$ and $\widehat{E}_6$ with negative levels, because \eqref{eq:positivity} was not satisfied there. However, in our cases, it turns out that \eqref{eq:positivity} and  $k+h^\vee>0$ are satisfied (we will show this explicitly below). Therefore, the Kazhdan-Lusztig formula reduces to the Kac-Wakimoto formula.}
To describe it, we first define $\Delta_{\lambda,+}^{re} = \{\alpha \in \Delta^{re}_+ | \langle \lambda,\alpha^\vee\rangle\in{\bf Z} \}$, where $\Delta_+^{re} = \Delta_+\setminus \{n\delta|n \in {\bf Z}^+\}$ is the set of real positive roots of $\mathfrak{g}$. Let us take simple roots, $\widehat{\alpha}_i$, of $\Delta_{\lambda,+}^{re}$ and define $W_\lambda$ as the Weyl group generated by simple reflections associated with $\widehat{\alpha}_i$. Note that $W_\lambda$ is a subgroup of the full Weyl group of $\mathfrak{g}$. For $w\in W_\lambda$, the shifted Weyl reflection of a weight $\lambda$ is defined by $w.\lambda \equiv w(\lambda + \rho) -\rho$, where $\rho$ is the Weyl vector of $\mathfrak{g}$. Now, suppose that the highest weight $\lambda$ satisfies
\begin{align}
\langle \lambda+\rho, \alpha^\vee\rangle >0
\label{eq:positivity}
\end{align}
for all $\alpha \in \Delta_{\lambda,+}^{re}$, and that the level, $k$, and the dual Coxeter number of $\mathfrak{g}$, $h^\vee$, satisfy $k + h^\vee > 0$. Then, it follows that \rcite{Kac-Wakimoto}
\begin{align}
\text{ch}\,L(\lambda) = \sum_{w\in W_\lambda}\epsilon(w)\; \text{ch}\, M(w.\lambda)~,
\label{eq:KW-formula}
\end{align}
where $\epsilon(w) = (-1)^k$ if $w$ is composed of $k$ simple Weyl reflections.\footnote{In Theorem 1 of \rcite{Kac-Wakimoto}, the highest weight $\lambda$ is supposed to be such that $\lambda + \rho\in K$ with $K$ being a special set of weights. For an affine Lie algebra associated with a finite dimensional simple Lie algebra, the set $K$ reduces to $\{\lambda \in \mathfrak{h}^*| \langle \lambda, \delta^\vee\rangle >0\}$. Then $\lambda + \rho\in K$ is equivalent to $0 < \langle \lambda + \rho, \delta^\vee\rangle = k + h^\vee$. } This means that $m_{\lambda,\mu} = \epsilon(w)$ if there exists $w\in W_\lambda$ such that $\mu = w.\lambda$, and otherwise $m_{\lambda,\mu}=0$. The expression \eqref{eq:KW-formula} is similar to the Weyl-Kac formula, but the sum is taken over $W_\lambda$ instead of the full Weyl group of $\mathfrak{g}$. Below, we apply this formula to the vacuum characters of the $\widehat{A}_1$ and $\widehat{A}_2$ algebras at level $k=-\frac{4}{3}$ and $k=-\frac{3}{2}$, respectively.

\subsection*{$\widehat{A}_1$ character with $k=-\frac{4}{3}$}

Suppose $\mathfrak{g} = \widehat{A}_1$ and $k=-\frac{4}{3}$. The finite part, $\bar{\mathfrak{g}} = A_1$, has a single simple root, $\alpha_1$, and therefore the set of positive roots of $\mathfrak{g}$ is $\Delta_+ = \{\alpha_1\}\cup \{ \pm \alpha_1 +n\delta| n\in{\bf Z}^+\}\cup \{n\delta| n\in{\bf Z}^+\}$. Its real part is then $\Delta_+^{re} = \{\alpha_1\}\cup \{\pm \alpha_1 + n\delta | n\in{\bf Z}^+\}$. The zeroth simple root is given by $\alpha_0 \equiv \delta - \alpha_1$. We want to evaluate the vacuum character of this algebra for the highest weight $\lambda = [-\frac{4}{3},0]$.\footnote{Here $[\lambda_0,\lambda_1]$ are the Dynkin labels of $\lambda$, namely, $\langle \lambda, \alpha_i^\vee \rangle = \lambda_i$. Note that $\sum_i \lambda_i = k$.} Since $\langle\lambda, (\pm \alpha_1 + n\delta)^\vee \rangle = - \frac{4n}{3}$, it follows that $\Delta_{\lambda,+}^{re} = \{\alpha_1\}\cup \{\pm \alpha_1 + 3n\delta | n\in{\bf Z}^+\}$. The simple roots of $\Delta_{\lambda,+}^{re}$ are $\alpha_1$ and $3\delta-\alpha_1 = 3\alpha_0  +2\alpha_1$, and therefore $W_\lambda$ is generated by $s_{\alpha_1}$ and $s_{3\alpha_0  +2\alpha_1}$, where $s_{\alpha}$ is the simple reflection associated with $\alpha$. Note that $\lambda = [-\frac{4}{3},0]$, $k=-\frac{4}{3}$, and $h^\vee = 2$ satisfy \eqref{eq:positivity} and $k+h^\vee>0$. Therefore, the character of $L(\lambda)$ is evaluated via \eqref{eq:KW-formula} as
\begin{align}\label{charA1A3}
e^{-\lambda}\text{ch}\, L(\lambda) = \frac{(1-z^2) - q^3(1-z^6)/z^2 + q^9(1-z^{10})/z^4 + \mathcal{O}(q^{18})}{\prod_{n=1}^\infty(1-q^n) (1-q^n z^{-2})(1-q^{n-1}z^2)}~,
\end{align}
where $q=e^{-\delta}$ and $z=e^{-\frac{1}{2}\alpha_1}$. Note here that $\frac{1}{2}\alpha_1$ is the fundamental weight of $su(2)$. In writing \eqref{charA1A3}, we normalize the character so that the highest weight contributes $1$.
By expanding this formula in powers of $q$, we obtain equation \eqref{eq:A1A3-index} and table \ref{table:A1A3-multiplicities}. 

\subsection*{$\widehat{A}_2$ character with $k=-\frac{3}{2}$}

Now suppose $\mathfrak{g} = \widehat{A}_2$ and $k=-\frac{3}{2}$. The finite part is $\bar{\mathfrak{g}} = A_2$, and its positive roots are $\bar{\Delta}_+ = \{\alpha_1,\alpha_2,\alpha_1+\alpha_2\}$, where $\alpha_1,\alpha_2$ are simple roots of $\bar{\mathfrak{g}}$. The real positive roots of $\mathfrak{g}$ are $\Delta_{+}^{re} = \overline{\Delta}_+\cup \{\pm \bar{\alpha} + n\delta |\bar{\alpha}\in\bar{\Delta}_+, n\in {\bf Z}^+\}$. The zero-th simple root is $\alpha_0 \equiv \delta - \alpha_1-\alpha_2$. We are interested in the highest weight module with highest weight $\lambda = [-\frac{3}{2},0,0]$. Since $\langle \lambda,(\pm\bar{\alpha}+n\delta)^\vee\rangle = -\frac{3n}{2}$, it turns out that $\Delta_{\lambda,+}^{re} = \overline{\Delta}_+\cup \{ \pm \bar{\alpha} + 2n\delta| \bar{\alpha}\in\overline{\Delta}_+, n\in{\bf Z}^+\}$. The simple roots of $\Delta_{\lambda,+}^{re}$ are $\alpha_1,\, \alpha_2$, and $2\delta -\alpha_1-\alpha_2 = 2\alpha_0 + \alpha_1+\alpha_2$, and therefore $W_\lambda$ is generated by $s_{\alpha_1},\,s_{\alpha_2}$, and $s_{2\alpha_0 + \alpha_1 + \alpha_2}$. Since $\lambda = [-\frac{3}{2},0,0]$, $k=-\frac{3}{2}$, and $h^\vee = 3$ satisfy \eqref{eq:positivity} and $k+h^\vee>0$, the character of $L(\lambda)$ is evaluated via \eqref{eq:KW-formula}.
We normalize the character so that the highest weight contributes $1$, and then write it in terms of $q=e^{-\delta},\, z_1=e^{-(\frac{2}{3}\alpha_1+\frac{1}{3}\alpha_2)}$ and $z_2=e^{-(\frac{1}{3}\alpha_1+\frac{2}{3}\alpha_2)}$. Note that $\frac{2}{3}\alpha_1+\frac{1}{3}\alpha_2$ and $\frac{1}{3}\alpha_1+\frac{2}{3}\alpha_2$ are the fundamental weights of $su(3)$. Expanding the normalized character in powers of $q$, we obtain equation \eqref{eq:A1D4-index} and table \ref{table:A1D4-multiplicities}.

\end{appendices}

\newpage
\bibliography{chetdocbib}

\end{document}